# The Behavior of Selected Diffuse Interstellar Bands with Molecular Fraction in Diffuse Atomic and Molecular Clouds

Haoyu Fan[1]
Key Laboratory of Optical Astronomy, National Astronomical Observatories, Chinese Academy of Sciences, Datun Road 20A, Beijing, 100012, China; hyfan@bao.ac.cn

Daniel E. Welty
Space Telescope Science Institute, 3700 San Martin Drive, Baltimore, MD, 21218, USA

Donald G. York[2]
Department of Astronomy and Astrophysics, The University of Chicago, 5640 S. Ellis Ave, ERC 577, Chicago, IL, 60615, USA; don@oddjob.uchicago.edu; 312 206 9950

Paule Sonnentrucker[3]
Space Telescope Science Institute, 3700 San Martin Drive, Baltimore, MD, 21218, USA

Julie A. Dahlstrom
Department of Physics and Astronomy, Carthage College, 2001 Alford Park Drive, Straz Center 94, Kenosha, WI, 53140, USA

Noah Baskes
Department of Astronomy and Astrophysics, The University of Chicago, 5640 S. Ellis Ave, ERC 577, Chicago, IL, 60615, USA

Scott D. Friedman
Space Telescope Science Institute, 3700 San Martin Drive, Baltimore, MD, 21218, USA

Lewis M. Hobbs
Yerkes Observatory, The University of Chicago, 373 W Geneva St, Williams Bay, WI, 53191, USA

Zihao Jiang
Department of Astronomy and Astrophysics, The University of Chicago, 5640 S. Ellis Ave, ERC 577, Chicago, IL, 60615, USA

Brian Rachford
Embry-Riddle Aeronautical University, 3700 Willow Creek Road, Prescott, AZ, 86303, USA

Theodore P. Snow
University of Colorado, 2000 Colorado Ave, Duane Physics Building, Rm. E226, Boulder, CO, 80309, USA

Reid Sherman
U.S. Global Change Research Program, 1800 G Street NW, Suite 9100, Washington, DC 20006, USA

Gang Zhao[1]
Key Laboratory of Optical Astronomy, National Astronomical Observatories, Chinese Academy of Sciences, Datun Road 20A, Beijing, 100012, China; gzao@nao.cas.cn

---

[1]University of Chinese Academy of Sciences, Beijing, 100049, China
[2]The Enrico Fermi Institute, University of Chicago, IL, USA
[3]European Space Agency

**Abstract:** We study the behavior of eight diffuse interstellar bands (DIBs) in different interstellar environments, as characterized by the fraction of hydrogen in molecular form ($f_{H2}$), with comparisons to the corresponding behavior of various known atomic and molecular species. The equivalent widths of the five "normal" DIBs ($\lambda\lambda$5780.5, 5797.1, 6196.0, 6283.8, and 6613.6), normalized to $E_{B-V}$, show a "Lambda-shaped" behavior: they increase at low $f_{H2}$, peak at $f_{H2} \sim 0.3$, and then decrease. The similarly normalized column densities of Ca, $Ca^+$, $Ti^+$, and $CH^+$ also decline for $f_{H2} > 0.3$. In contrast, the normalized column densities of Na, K, CH, CN, and CO increase monotonically with $f_{H2}$, and the trends exhibited by the three $C_2$ DIBs ($\lambda\lambda$4726.8, 4963.9, and 4984.8) lie between those two general behaviors. These trends with $f_{H2}$ are accompanied by cosmic scatter, the dispersion at any given $f_{H2}$ being significantly larger than the individual errors of measurement. The Lambda-shaped trends suggest the balance between creation and destruction of the DIB carriers differs dramatically between diffuse atomic and diffuse molecular clouds; additional processes besides ionization and shielding are needed to explain those observed trends. Except for several special cases, the highest $W_\lambda(5780)/W_\lambda(5797)$ ratios, characterizing the so-called "sigma-zeta effect", occur only at $f_{H2} < 0.2$. We propose a sequence of DIBs based on trends in their pair-wise strength ratios with increasing $f_{H2}$. In order of increasing environmental density, we find the $\lambda$6283.8 and $\lambda$5780.5 DIBs, the $\lambda$6196.0 DIB, the $\lambda$6613.6 DIB, the $\lambda$5797.1 DIB, and the $C_2$ DIBs.

## 1. INTRODUCTION

Diffuse interstellar bands (DIBs) were first observed in stellar spectra in 1919 (Heger 1922; Herbig 1995; McCall & Griffin 2011), though an archive plate of HD 80077 examined by A. J. Cannon for the Henry Draper Catalogue likely indicates the presence of the strong broad $\lambda$4428.8[4] DIB several years earlier (Code 1958; Oka & McCall 2011). The general idea that DIBs could be due to interstellar molecules goes back to the 1930s (e.g., Merrill 1934, 1936; Swings 1937; Swings & Rosenfeld 1937; Douglas & Herzberg 1941). After nearly 100 years of observations, more than 500 such diffuse, as yet unidentified absorption features have been cataloged in the optical spectra of background stars (Hobbs et al. 2008, 2009). The most distinctive property of the DIB profiles is their width, which ranges from FWHM ~20 km s$^{-1}$ for the narrow $\lambda$6196.0 DIB to more than 1000 km s$^{-1}$ for the broad $\lambda$4428.8 DIB. The DIBs are thus much broader than the lines due to known atomic species and the resolved bands due to known molecular species, which can have widths as small as 0.5 km s$^{-1}$ for individual velocity components (e.g. Welty et al. 1994; Crane et al. 1995; Welty & Hobbs 2001).

The substructure visible in very high resolution spectra of some DIBs in "simple" sight lines strongly suggests that the carriers are molecular (e.g. Sarre et al. 1995; Kerr et al. 1998; Galazutdinov et al. 2008). A number of carbon-based molecules have been proposed as DIB carriers, including polycyclic aromatic hydrocarbons (PAHs), carbon chain molecules, and more complicated structures such as fullerenes (e.g. Foing & Eherenfreund 1994; Gredel et al. 2011; Herbig 1995 for a review). Some of those molecules have also been suggested to give rise to unidentified features seen in near-IR emission (e.g., Rouan et al. 1997, Ruiterkamp et al. 2005, Maier et al. 2004; Kwok & Zhang 2013). On the other hand, recent detailed analyses of DIB profiles have suggested that several of the stronger DIBs may be due to small molecules ($\leq$ 7 heavy atoms) (Oka et al. 2013; Huang & Oka 2015). While a number of such small molecules have been considered over the years, most detailed comparisons between laboratory and astronomical spectra have not yielded adequate matches (e.g., Tulej et al. 1998; Motylewski et al. 2000; Lakin et al. 2000; McCall et al. 2001; Guthe et al. 2001; Salama et al. 2011). Some five near-IR DIBs were recently attributed to the fullerene cation $C_{60}^+$ (Campbell et al. 2015; Walker et al. 2015; though see Galazutdinov et al. 2017).

Along with laboratory work, correlation studies have been performed to investigate the nature of carriers of the DIBs, by comparing DIB equivalent widths (EWs) with different interstellar parameters such as $E_{B-V}$, the column densities of interstellar species ($N$(X)), and the strengths of other DIBs (e.g. Herbig 1993; Cami et al. 1997; Friedman et al. 2011; Vos et al. 2011). These studies led to the groupings of some DIBs into classes and families, which may reflect common properties of the carriers within each group (e.g. Krełowski & Walker 1987; Josafatsson & Snow 1987; Westerlund & Krełowski 1989; Krełowski et al. 1997; Cami et al. 1997; Weselak et al. 2001). One such family is the so-called "$C_2$ DIBs", whose normalized EWs, $W_\lambda(X)/W_\lambda(6196)$, grow rapidly when $C_2$ molecules become more abundant in the sight line, although their carriers are not necessarily chemically related to the $C_2$ molecules (Thorburn et al. 2003).

---

[4]Wavelengths used here are from Table 2 of Hobbs et al. (2008). These values can differ between authors due to differences in the definition of the central wavelength (e.g., un-weighted or weighted).

Studies have revealed good correlations between some strong DIBs, with a nearly perfect correlation between the λ6196.0 and λ6613.6 DIBs (Cami et al. 1997; Moutou et al. 1999; Galazutdinov et al. 2002; McCall et al. 2010). However, significant sight line to sight line differences in the relative strengths of some DIBs have also been noted (e.g., Hobbs et al. 2008, 2009). The best known example involves the strength ratio of the λ5780.5 and λ5797.1 DIBs, which can vary by an order of magnitude among different sight lines (Krełowski & Walker 1987; Krełowski & Westerlund 1988; Vos et al. 2011; Kos & Zwitter 2013; Welty et al. 2014). This behavior is known as the sigma-zeta effect, named after two early examples, σ Sco (HD 147165) and ζ Oph (HD 149757), whose $W_\lambda(5780)/W_\lambda(5797)$ ratios differ by a factor of 2.2. These two sight lines appear to probe rather different interstellar environments. For example, the σ Sco sight line has a flat UV extinction curve and low abundances of diatomic molecular species, while the ζ Oph sight line has a steeper than average UV extinction curve and much higher molecular abundances (e.g., Morton 1975; Voshchinnikov et al. 2006). Such observations point to the possibility that DIBs can be used as indicators of different interstellar environments in the sight line.

Since 1999, we have obtained moderately high resolution, high signal-to-noise ratio (SNR) spectra for 421 sight lines sampling a wide range of ISM environments, with $E_{B-V}$ ranging from 0.00 to 3.31 mag. These spectra, together with the DIB measurements and various ISM parameters, are being placed in a public online database[5]. We selected spectra from this database to produce a new set of uniform measurements, independent of published values, to undertake a systematic study of the strengths of eight "representative" DIBs and various other interstellar parameters, to better understand the behavior of those DIBs in different interstellar environments.

This work is organized as follows. Section 2 includes a description of the stars and DIBs selected, and the measuring techniques adopted to ensure a uniform set of measurements. In Section 3, we present the behavior of $W(DIB)/E_{B-V}$ and the $W_\lambda(5780)/W_\lambda(5797)$ ratio in different interstellar environments. These behaviors are discussed and compared to better known constituents of the ISM in Section 4, which concludes with a proposed ordering of the DIBs, based on their behavior in the diffuse clouds. Finally, Section 5 summarizes the conclusions reached in this work. Appendix A explores several different ways to define and measure the λ5797.1 DIB. Appendix B includes graphical presentations of the strength of the DIBs studied here.

## 2. OBSERVATIONS
2.1 Optical Spectra

For this exploration of the environmental dependence of the DIBs, 186 sight lines, with $E_{B-V}$ ranging from 0.00 to 3.31 mag, $f_{H2}$ ranging from less than $10^{-4}$ to 0.83, and sampling a variety of regions in the Galactic ISM within several kpc of the Sun, were selected from our larger set of observations. As will be discussed below, the very different environmental conditions found along these 186 sight lines affect the behavior of both the known atomic and molecular species and the DIB carriers.

---

[5] see http://dib.uchicago.edu/

Most of the spectra were obtained with the ARC echelle spectrograph (ARCES) on the 3.5m telescope at Apache Point Observatory (APO) and many of these spectra were used in the series of papers "Studies of Diffuse Interstellar Bands" (Thorburn et al. 2003; Hobbs et al. 2008; Hobbs et al. 2009; McCall et al. 2010; Friedman et al. 2011; and related papers:  McCall et al. 2001; Dahlstrom et al. 2013; Welty et al. 2014).  As several of those papers include detailed characterizations of the ARCES spectra and descriptions of the standard data reduction procedures, only a brief summary will be given here.

The ARCES spectra cover a wavelength region between 3500Å and 11000Å, and the resolving power is 33,000 with a 1.6" slit, corresponding to a velocity resolution of 9 km s$^{-1}$ (Wang et al. 2003).   For each sight line, multiple exposures were combined to achieve a nominal SNR of ~1,000 (per resolution element) around 6400Å.  Lightly reddened standard stars, covering similar ranges in spectral type and luminosity class as our target sample and (ideally) with relatively low projected rotational velocity, were observed to similar SNR in order to enable recognition of any stellar lines blended with the DIBs.  Cosmic ray removal, bias subtraction, and flat fielding were done with standard techniques, and telluric lines were corrected based on humidity and air mass.  The wavelength scale is set by defining the strongest component of the interstellar K I line at laboratory wavelength 7698.9645Å as zero velocity (Morton 2003).  If interstellar K I is not detected, the analogous procedure is followed using the Na I line.  The reduction of the raw spectra was done by J. D.

Eight of our stars[6] were observed with the Magellan Inamori Kyocera Echelle (MIKE), a double echelle spectrograph on the Clay Telescope at Las Campanas Observatory, Chile, with wavelength coverage between 3200Å and 10,000Å (Bernstein et al. 2003).  The effective resolving power is about 45,600 in the blue and 36,000 in the red for these spectra.  Another five spectra[7] are from the archive of the Fiberfed Extended Range Optical Spectrograph (FEROS) mounted on the ESO 1.52m telescope at La Silla observatory, Chile (Kaufer et al. 2000), which has a resolving power of 48,000 and wavelength coverage between 3560Å and 9200Å.  Both spectrographs thus have similar resolving power and wavelength coverage to ARCES.  The spectral extractions and stacking of multiple spectra from the MIKE and FEROS were performed using an IRAF-based pipeline constructed by Z. J.

2.2 DIBs Chosen for Investigation

Of the more than 500 DIBs recognized between 3500Å and 8500Å (Hobbs et al. 2008, 2009), eight were selected for this study.   These eight DIBs are generally relatively strong (so they can be measured even in lightly reddened sight lines), fairly well separable from any other nearby DIBs (with caveats noted below), and have minimal contamination from telluric and stellar lines. Five of the eight DIBs ($\lambda\lambda$5780.5, 5797.1, 6196.0, 6283.8, and 6613.6) have been examined in a number of previous investigations.   While the prior studies have provided evidence of differences in behavior among those five well known DIBs, we will nonetheless refer to them as "normal" DIBs.  That designation is in contrast to the other three DIBs ($\lambda\lambda$4726.8, 4963.9, 4984.8) studied in this paper that belong to the class of "$C_2$ DIBs" (Thorburn et al. 2003), which as far as we can tell includes less than 10% of the currently known DIBs (see D. G. York et al., in preparation).  As will be seen below, the behavior of the three $C_2$ DIBs is in sharp contrast to that of the normal DIBs, particularly when denser regions with higher molecular fractions are present along the sight line.

---

[6]HD62542, HD94414, HD108927, HD114213, HD147701, HD156738, HD157246, and BD -14 5037

[7]HD110432, HD112244, HD152234, HD152235 and HD152236

2.3 Measurement of DIB Equivalent Widths
2.3.1 Choice of Integrated Equivalent Widths

Spectra obtained at moderately high to high resolution indicate that the profiles of the DIBs generally are not Gaussian, and that they are not necessarily uniform in width or shape in different sight lines (e.g. Sarre et al. 1995; Galazutdinov et al. 2008; Hobbs et al. 2008, 2009).  In some sight lines, those differences likely reflect the underlying complex component structure as discernible, for example, in the corresponding absorption lines from Na, K, and $Ca^+$ (Welty et al. 1994, 1996, 2001).  In other cases, most notably toward the heavily reddened star Herschel 36, local physical conditions can modify the DIB profiles (Dahlstrom et al. 2013; Oka et al. 2013). While the very strong, extended redward wings seen for some of the DIBs toward Herschel 36 are (so far) unique to that sight line, weaker such wings and other, more subtle differences are seen in some other cases.  Moreover, as long as the DIB carriers remain unidentified, the intrinsic profiles of the corresponding DIBs (and their potential variations in different physical environments) remain unknown, hindering attempts to identify and separate the contributions from blended DIBs and/or to determine template DIB profiles appropriate for all sight lines. Given those considerations, and given the resolution and SNR characterizing our spectra, we therefore have measured the EWs of the DIBs by direct integration over the absorption-line profiles in each case without prior assumptions as to the shapes or widths of the DIB profiles. All EW measurements were performed by F. H., using the semi-automated program described below.

2.3.2 Continuum Issues

Reliable, uniform measurement of the DIB EWs depends on accurate determination of both the local continuum (e.g., taking into account the effects of any stellar absorption features, which can vary with spectral type) and the interval over which the DIB absorption is to be integrated.  To illustrate some of these issues, Figure 1 shows 24Å-wide spectral segments around the eight DIBs chosen for this study, for the two DIB atlas stars HD183143 (Hobbs et al. 2009) and HD204827 (Hobbs et al. 2008).  In the figure, tick marks give the positions of DIBs identified in the two sight lines, arrows show the locations of stellar absorption lines seen in the respective unreddened comparison stars, and the solid dots indicate typical integration limits (appropriate for most of the sight lines in our sample) for the DIBs.  Four of the eight DIBs ($\lambda\lambda 4726.8$[8], 4984.8, 6196.0, and 6613.6) are relatively narrow and well isolated from other DIBs or significant telluric absorption.  They are generally not blended with stellar lines either, which makes their measurement straightforward.  The narrow $\lambda 4963.9$ $C_2$ DIB can be slightly blended with a much weaker DIB at 4965.2 Å, but the two are generally easily separable.

Measurement of the strong DIBs $\lambda\lambda 5780.5$, 5797.1, and 6283.8, however, requires more consideration:

---

[8]The profile of $\lambda 4726.8$ suggests this DIB is a blend of two DIBs, but in this study we measure the entire feature as one DIB.  Detailed comments are given later in the paper in Section 3.5.

1) The $\lambda$5780.5 DIB is the dominant feature in a complex of DIB absorption which contains a broader (~20Å wide), shallow component and a number of weaker, relatively narrow components (Hobbs et al. 2008, 2009; Dahlstrom et al. 2013; and references therein).  As in most previous studies (e.g., Galazutdinov et al. 2004; Friedman et al. 2011), we focus on the dominant narrow feature at 5780.5 Å (see the integration limits in Fig. 1).  Because of possible significant stellar contamination of the $\lambda$5780.5 DIB for stellar types B4 and later, those measurements are excluded from our correlation analyses of this DIB.

2) The strong, broad absorption typically assigned to the $\lambda$6283.8 DIB can include contributions from several weak, relatively narrow DIBs with different relative strengths in different sight lines. Consistent with most previous work, we measure the EW of the entire complex.  This DIB sits on a telluric band and signs of the residuals from under or over correction can be seen in some of our spectra.  The uncertainties arising from these telluric residuals are not included in the errors of measurements of the $\lambda$6283.8 DIB, but flags are assigned to measurements made in spectra where the telluric correction is not satisfactory (see Table 1).

3) Different measurement conventions for the $\lambda$5797.1 DIB have been adopted in previous studies, reflecting assumptions concerning two much weaker absorption features at 5795.2 and 5793.2Å.  Thorburn et al. (2003) and Friedman et al. (2011), for example, included the $\lambda$5795.2 DIB as part of the $\lambda$5797.1 DIB and measured them together, while Galazutdinov et al. (2004) excluded the $\lambda$5795.2 DIB by using a lower continuum (essentially assuming the $\lambda$5795.2 DIB to be part of a broader absorption feature).  After detailed examination of various options (see Appendix A), we have chosen a third method for measuring the $\lambda$5797.1 DIB, in which the continuum is defined as in Friedman et al. (2011) but the integration limit at the short wavelength side of the DIB is placed at the local maximum between the $\lambda$5795.2 and the $\lambda$5797.1 DIBs. This choice essentially assumes that the $\lambda$5795.2 DIB is relatively narrow.  Comparisons among the three methods indicate that the specific choice does not significantly affect the conclusions drawn from our correlation analyses, as long as the same method is used consistently for all sight lines, essentially because the EW is dominated in all three cases by the strong $\lambda$5797.1 feature (Appendix A).  An additional complication, for some O stars with high $v\sin i$, is that the redward wing of the $\lambda$5797.1 DIB can be blended with a stellar C IV line.  In many of those cases, an adequate correction for the stellar line could be performed by dividing the observed spectrum by a symmetric Gaussian profile fitted to the stellar line.  Sight lines where such a correction was not adequate[9] (e.g., for spectroscopic binaries observed at multiple epochs, where the stellar line profiles in the summed spectra are asymmetric), were excluded from the correlation analyses involving the $\lambda$5797.1 DIB.

2.3.3 Semi-Automated Measuring Procedure

The DIBs in each sight line were measured using a semi-automated routine (arcexam; written by D. E. W.).  For each DIB to be measured, the program first displays a small section of the spectrum around that DIB and provides an initial fit to the local continuum. Corresponding spectral segments from the two DIB atlas stars (Hobbs et al. 2008, 2009) and a telluric reference star (10 Lac, uncorrected for telluric absorption) are included to provide guidance regarding the expected DIB widths and profiles (as seen toward the atlas stars) and the locations of possible residuals from the correction for telluric absorption.   For this study of eight DIBs, spectra of several lightly reddened, low-$v$sin$i$ stars of each spectral type were examined to identify possible blends of each DIB with stellar absorption lines.   The user can either accept the program's continuum fit or make adjustments.

---

[9]These stars are HD 54662, 57060, 57061, 110432, 93521, 147701 and 210839.

Once the continuum is determined, the program determines the extent of the absorption (i.e., the integration limits), then measures the EW as well as the first four moments of the absorption-line profile (central wavelength, width, skewness, and kurtosis).   Again, the user can either accept the program's choice of integration limits or choose different ones.   Uncertainties on the EWs (1-σ) and limits (3-σ) for undetected DIBs are based on the fluctuations in the local continuum and the width of the DIB in the two atlas stars (Hobbs et al. 2008, 2009); for these relatively broad features, the uncertainties are often dominated by uncertainties in the placement of the continuum, rather than by photon noise alone (e.g., Sembach & Savage 1992). Depending on such factors as SNR, spectral type, and $v$sin$i$, the program typically gives acceptable measurements for over 90% of the isolated DIBs without user intervention, and manual fine adjustment of the continuum level and/or integration limits can usually settle the rest. Any possible contamination from telluric residuals, stellar lines, or other DIBs is flagged, and those measurements are excluded from the correlation analyses (except for the λ6283.8 DIB, discussed above).   Table 1 lists the star name; the spectral type; the color excess, $E_{B-V}$; the column densities of neutral hydrogen and molecular hydrogen (measured or estimated, see Section 3.1), $N$(H) and $N$(H$_2$); the fraction of molecular hydrogen (by mass), $f_{H2}$; and the measured EWs of the eight DIBs we consider for the 186 sight lines in this survey.

Friedman et al. (2011) studied the correlations between the EWs of eight normal DIBs ($\lambda\lambda$5487.7, 5705.1, 5780.5, 5797.1, 6196.0, 6204.5, 6283.8, and 6613.6), $N$(H), $N$(H$_2$), and $E_{B-V}$, for 133 sight lines observed with ARCES.   The present study has 104 stars and four DIBs in common with those of Friedman et al. (DIBs $\lambda\lambda$5780.5, 6196.0, 6283.8, and 6613.6) for which the choice of continuum placement and integration limits are very similar.   The fifth DIB in common, $\lambda$5797.1, is discussed above.   Figure 2 compares the measurements of DIBs $\lambda\lambda$5780.5 and 6613.6 obtained in the two studies.   Both correlations are essentially perfect (correlation coefficient $r$ = 0.999, slopes are consistent with unity, intercepts are consistent with zero), given the errors in the measurements (e.g., from continuum placement, subtle blends of adjacent or non-obvious DIBs; see Hobbs et al. 2008 and McCall et al. 2010).   The semi-automated program used in this study thus appears to yield results entirely consistent with those obtained by hand, but takes much less time -- making the uniform measurement of larger sets of DIBs in even more sight lines quite feasible.

2.4 The Reddening, $E_{B-V}$

The $E_{B-V}$ values were compiled by L. M. H. from standard sources (see Friedman et al. 2011). The observational data, namely spectral type and observed (B-V), used here for determining the needed stellar properties were taken, as available, from one of the following sources, and generally in order of preference as follows:
1) The Bright Star Catalogue or its Supplement (BSC);
2) The Hipparcos Input Catalog (HIC);
3) The various measurements collected for a given star in the SIMBAD database;
4) The general literature.

If data were available in neither the BSC nor the HIC, informed choices were sometimes necessary among various distinct data sets available in SIMBAD or the literature. As a last resort in the rare cases in which the preferred source gives a spectral type and/or (B-V) significantly different from the values found in one or more of the subsequent sources, values from the latter were adopted after an investigation of those differences.

Estimated values of intrinsic color $(B-V)_0$ and of absolute magnitude $M_v$ are required in order to calculate $E_{B-V}$ and, in the absence of a reliable trigonometric parallax, a stellar distance. The calibrations we adopted for both $(B-V)_0$ and $M_v$ in terms of spectral type are uniformly those of Johnson (1963). Our first requirement in adopting these particular calibrations was methodological consistency, i.e. the need to avoid any systematic errors introduced by modifying our choices of these calibrations during the course of a project which has now spanned 20 years. Over those decades, the various newer $(B-V)_0$ calibrations available have generally differed modestly from Johnson's, especially when the differences among the measured (B-V) colors and among the spectral types that were adopted by different authors to establish the respective calibrations are considered. The corresponding differences between the various newer $M_v$ calibrations and that of Johnson are sometimes somewhat larger. Since the derived stellar distances are of minimal direct interest in this paper, and since reliable trigonometric distances are directly available for some of our program stars, we have again consistently used Johnson's calibration where required, to maintain internal consistency. We adopt the value 0.03 mag for the typical error on $E_{B-V}$ as a fair representation that covers the major uncertainties: the direct errors of observation of the colors, errors in the spectral type necessary to derive the intrinsic colors of each star, and errors in the calibration to convert from a nominal spectral type to intrinsic colors. The second uncertainty is often likely to be the major contributor, for the star sample we are using.

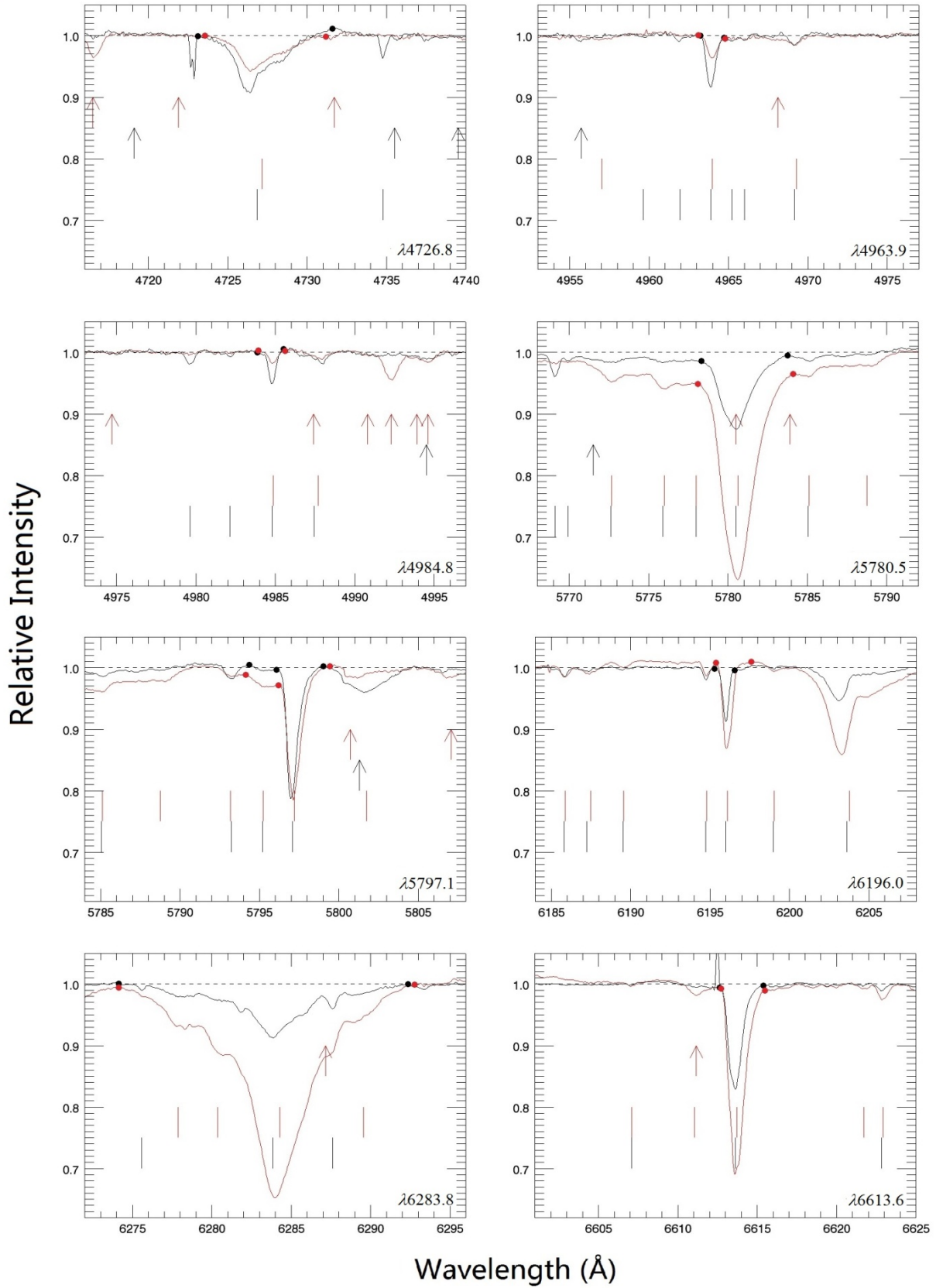

Figure 1. Measurements of the eight DIBs of interest in this study from two well-observed sight lines: HD 183143 (red) and HD 204827 (black). The arrows mark stellar lines noted in beta Ori (B8Iae, a standard star of low extinction for HD 183143, B7Iae), and 10 Lac (O9V, standard for HD 204827, B0V).   Vertical bars are nearby DIBs from Hobbs et al. (2008, 2009).   The colors of the bars and arrows correspond to the colors of the spectral plots.   The dots demonstrate the integration limits, which are chosen by where the features reach the continuum rather than at fixed wavelengths.   Multiple dots are given for the $\lambda$5797.1 DIB to represent different measuring techniques.   Note that both stars have similar $E_{B-V}$, but four of the eight DIBs are stronger in HD 183143 (DIBs $\lambda\lambda$5780.5, 6196.0, 6283.8, and 6613.6), while the $\lambda$5797.1 DIB is comparable, and the $C_2$ DIBs are much stronger in the sight line of HD 204827.   The double absorption seen to the left of the $\lambda$4726.8 DIB and the spike next to the $\lambda$6613.6 DIB are artifacts.

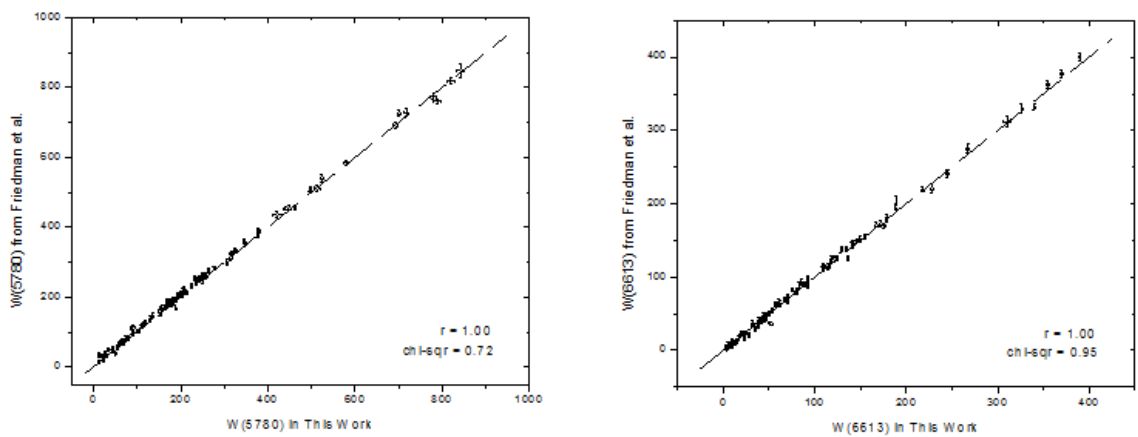

Figure 2. Comparison between $W_\lambda$(5780) and $W_\lambda$(6613) measured in this work and by Friedman et al. (2011).   The dotted lines indicate $Y = X$.   The two sets of data match essentially perfectly in both cases, with Pearson linear correlation coefficients ($r$) greater than 0.99, the slopes of the relationships consistent with unity, and the intercepts consistent with zero -- indicating no systematic difference or offset.   The values for chi-squared for both correlations are also close to unity, and we conclude that the uncertainties of the measurements are properly estimated. The calculations were done as described in Section 3.5.

Table 1: Spectral Types and Values of $E_{B-V}$, $N(H)$, $N(H_2)$, $f_{H2}$, and EWs of the Eight DIBs for 186 O and B Stars[a]

| HD | Spectral Type | $E_{B-V}$ (mag) | $Log_{10}N(H)$ | $Log_{10}N(H_2)$ | $f_{H2}$ | $W_\lambda(4726)$ (mÅ) | $W_\lambda(4963)$ (mÅ) | $W_\lambda(4984)$ (mÅ) | $W_\lambda(5780)$ (mÅ) | $W_\lambda(5797)$ (mÅ) | $W_\lambda(6196)$ (mÅ) | $W_\lambda(6283)$ (mÅ) | $W_\lambda(6613)$ (mÅ) |
|---|---|---|---|---|---|---|---|---|---|---|---|---|---|
| 886 | B2IV | 0.01±0.01 | | | | | <0.4 | <4.3 | 4.5±0.7 | 0.9±0.2 | | | <1.8[c] |
| 1544 | B0.5III | 0.43±0.03 | | | | 65.6±3.0 | 4.8±0.4 | 3.2±0.6 | 388.3±2.5 | 91.1±1.1 | 39.4±0.4 | 807.7±11.2 | 162.1±1.6 |
| 2905 | B1Iae | 0.33±0.05 | 21.28±0.09 | 20.27±0.14 | 0.2±0.1 | 49.6±1.8[c] | 6.3±0.2 | 4.6±0.5 | 309.9±1.7 | 89.7±0.8 | 27.7±0.2 | 596.3±5.5 | 125.0±1.2 |
| 10516 | B2Vep | 0.20±0.05 | <20.54 | 19.08±0.14 | >0.065 | 17.1±1.5[c] | <1.1[c] | 1.8±0.3 | 58.6±1.1 | 9.8±0.5 | 5.7±0.3 | 123.2±2.6 | 10.5±0.7[c] |
| 11415 | B3III | 0.05±0.02 | 20.53±0.10[b] | <19.32[b] | <0.11 | 26.1±2.2[c] | 1.0±0.4[c] | <1.0 | 29.9±1.7 | 4.8±0.5 | | 35.2±2.8 | 4.2±0.5 |
| 13267 | B5Iae | 0.42±0.03 | | | | 63.7±2.9 | 9.3±0.5 | 3.2±0.4 | 279.2±1.1[c] | 78.0±1.2 | 26.4±0.5 | 818.8±14.4 | 96.3±1.1 |
| D +56 508 | B2III | 0.53±0.03 | | | | 144.7±4.5[c] | 21.0±0.4[c] | 4.2±0.2 | 363.2±5.3 | 108.1±0.7 | 27.7±0.4 | 935.1±12.5 | 144.2±0.8c |
| 14134 | B3Ia | 0.58±0.03 | | | | 109.9±2.6 | 16.6±0.3 | 5.5±0.7 | 307.3±1.6 | 95.8±0.6 | 30.5±0.3 | 819.2±13.5 | 126.9±0.8 |
| 14143 | B2Ia | 0.67±0.03 | | | | 134.7±2.6 | 25.8±1.1[c] | 4.1±0.4 | 323.9±3.2 | 116.7±0.8 | 37.4±0.3 | 894.8±15.6 | 129.2±1.2 |
| 14633 | O8.5V | 0.11±0.03 | 20.60±0.06 | <19.11±0.00 | <0.061 | 18.7±1.8 | <1.6 | <1.6 | 39.4±1.8 | 11.6±1.1[d] | 5.1±0.2 | 68.9±7.6 | 15.2±1.4 |
| 14818 | B2Iae | 0.28±0.03 | | | | 81.9±2.7 | 11.7±0.5[c] | 4.6±0.4 | 308.0±3.3 | 91.1±1.9 | 30.1±0.4 | 907.5±30.7 | 127.0±1.2 |
| 15558 | O5e | 0.83±0.08 | 21.52±0.18 | 20.89±0.15 | 0.3±0.2 | 118.9±4.1 | 14.0±0.6 | 5.7±0.5 | 500.5±5.3 | 194.9±0.6 | 60.7±0.7 | 1097.8±17.7 | 237.1±1.0 |
| 18326 | O7V | 0.70±0.03 | | | | 100.7±3.6 | 13.0±0.5 | 6.3±1.5 | 399.4±4.1 | 145.4±3.3[d] | 45.2±0.6 | 996.8±14.8 | 183.0±1.4 |
| 19374 | B1.5V | 0.13±0.03 | 21.04±0.10[b] | 19.64±0.13[b] | 0.07±0.04 | 60.5±2.2[c] | 5.5±0.3 | 0.7±0.2 | 130.1±2.4 | 38.0±1.1 | 13.9±0.5 | 314.8±9.6 | 39.8±1.1[c] |
| 19820 | O9IV | 0.79±0.03 | 21.45±0.10[b] | 20.91±0.13[b] | 0.37±0.12 | 92.2±1.4 | 14.5±0.4 | 6.3±0.2 | 416.0±1.7 | 124.9±1.4[d] | 46.7±0.3 | 1052.0±12.0 | 209.9±0.8 |
| 21483 | B3III | 0.56±0.04 | 21.00±0.08 | 20.96±0.13[b] | 0.65±0.11 | 109.4±3.7[c] | 16.6±0.3[c] | 9.6±0.6 | 169.9±1.9 | 89.9±1.1 | 20.3±0.3 | 340.2±5.7 | 92.9±0.9 |
| 22951 | B0.5V | 0.27±0.03 | 21.04±0.13 | 20.46±0.14 | 0.3±0.1 | 47.8±3.0 | 8.2±0.4 | 2.0±0.4 | 110.3±2.0 | 45.4±1.6 | 14.1±0.5 | 208.0±8.2 | 47.8±0.7 |
| 23180 | B1III | 0.31±0.03 | 20.82±0.09 | 20.60±0.12 | 0.5±0.1 | 78.4±2.3[c] | 11.6±0.9[c] | 7.9±0.6 | 89.5±1.4 | 69.7±0.4 | 14.0±0.4 | 183.3±7.9 | 50.3±0.4[c] |
| 281159 | B5V | 0.85±0.05 | 21.39±0.30 | 21.09±0.19 | 0.50±0.28 | 111.8±3.9 | 23.9±0.8 | 11.7±0.6 | 310.2±1.8[c] | 110.5±1.3 | 31.2±0.3 | 582.2±8.4 | 146.5±1.4 |
| 24398 | B1Ib | 0.31±0.03 | 20.80±0.08 | 20.67±0.14 | 0.6±0.1 | 77.1±3.4[c] | 6.4±0.6 | 3.8±0.2 | 90.5±4.8 | 62.1±0.9 | 17.9±0.3 | 169.8±7.2 | 60.8±0.7 |

[a] The full table is available as a machine readable file in the online version.
[b] Column density calculated using surrogates (see Section 3.1).
[c] Possible stellar contamination.
[d] The profile of the $\lambda$5797.1 DIB corrected from stellar contamination.

## 3. RESULTS

3.1 Molecular Fraction of Hydrogen, $f_{H2}$

The integrated molecular fractions of hydrogen for different sight lines are compiled in our database of DIB stars (Friedman et al. 2011). We have attempted to include all stars with $H_2$ detections from the *Copernicus* (Savage et al. 1977; Bohlin et al. 1978) and *FUSE* satellites (for example, Rachford et al. 2002, 2009), and 108 stars in this work have direct measurement of $f_{H2}$ = $2N(H_2)/[N(H)+2N(H_2)]$. When direct measurements are not available, the EW of the $\lambda$5780.5 DIB is used as a surrogate for atomic hydrogen (28 sight lines; Herbig 1993; Friedman et al. 2011; see Figure 3A in this paper), and $N(H_2)$ is estimated based on $N(CH)$ (33 sight lines; e.g. Sheffer et al. 2008; see Figure 3B in this paper). The use of surrogates gives us 36 more $f_{H2}$ values, for a total of 144.

Concerning the use of surrogates for $N(H)$ and $N(H_2)$, we note there are some sight lines that do not follow the general trends. The outliers more than 3-σ away from the best-fit lines are highlighted in red in Figure 3. Most of the sight lines with weaker than 'expected' $W_\lambda(5780)$ versus $N(H)$ appear to have stronger than average radiation fields, while sight lines with stronger than 'expected' $N(CH)$ versus $N(H_2)$ may include regions where CH is formed largely by non-thermal effects (e.g., Federman et al. 1985; Zsargó & Federman 2003; Godard et al. 2014). As such cases are very few in number, they have little effect on mean relationships. In particular, the scatter in all figures related to $f_{H2}$ in this paper (see Figure 4 and beyond) is not much different for the blue points (indicating use of surrogate $f_{H2}$ values) and for the black points (using direct $f_{H2}$ measurements).

3.2 Behavior of DIBs at Different $f_{H2}$

In this study, we use the fraction of hydrogen in molecular form, $f_{H2}$, as an indicator of the 'typical' local hydrogen density in the main interstellar clouds in each line of sight (e.g., Cardelli 1994). The integrated $f_{H2}$ values for each sight line are plotted versus the normalized DIB EWs, in Figure 4. The color excess, a rough measure of the total amount of interstellar material, ranges from 0.00 to 3.31 mag in our data sample, and we use $E_{B-V}$ to normalize the EWs of the DIBs as $W(DIB)/E_{B-V}$. This represents an attempt to get beyond the general tendency for all interstellar tracers to increase with $N(H_{tot})$, so that more local environmental effects in the dominant interstellar clouds might be discernible. Given the large relative uncertainties when $E_{B-V}$ is small, we omit from the correlation analyses any measurements from sight lines whose reddening is smaller than 0.1 mag. A similar plot, where the EWs of the DIBs are not normalized, is provided in Appendix B. We note that the general behavior of the DIBs does not change upon normalization, although the normalization does greatly reduce scatter due to differences in distance and reddening, especially for $f_{H2} > \sim 0.2$ (Figure B1).

Twelve of the sight lines are highlighted in Figure 4 and in the plots following for their well-studied history or interesting behavior. HD 183143 (○) has often been used as a "standard" sight line for DIB studies (e.g. Herbig 1995; Hobbs et al. 2009). HD 147165 (σ Sco, ＋) and HD 149757 (ζ Oph, X) represent the defining sight lines of the sigma-zeta effect. The strongest $C_2$ DIB detections are made in the sight lines toward NGC2024-1 (●), HD 204827 (★), BD -14 5037 (■), and Cyg OB2 #5 (▲). HD 37903 (△) at $f_{H2}$ = 0.52, HD 73882 (□) at $f_{H2}$ = 0.67, and HD 200775 (✳) at $f_{H2}$ = 0.83 are the most significant outliers in Figure 5 (see section 3.3). For HD 37061 (◇) and perhaps for HD 62542 (◆) (Cardelli & Savage 1988) intense radiation fields may weaken the DIBs. Finally, some of the DIBs toward Herschel 36 (☆) exhibit an anomalously strong extended tail towards the red (ETR), likely due to a strong local IR field (Dahlstrom et al. 2013; Oka et al. 2013).

The normalized EWs of all five normal DIBs have a Lambda-shaped behavior with respect to $f_{H2}$ (Figure 4D through 4H). This behavior is also found for three additional DIBs ($\lambda\lambda5487.7$, 5705.1, and 6204.5) from Friedman et al. (2011). By "Lambda-shaped behavior" we mean $W(DIB)/E_{B-V}$ increases with $f_{H2}$ at low $f_{H2}$, reaches a peak at $f_{H2} \sim 0.3$, and then declines with increasing $f_{H2}$ thereafter. This is a confirmation that $W(DIB)/E_{B-V}$ for normal DIBs can be weaker in sight lines containing dense cloud regions (e.g., Wampler 1966; Adamson et al. 1991). The three $C_2$ DIBs, on the other hand, show similar growth when $f_{H2}$ is small, but the peak and declining part at larger $f_{H2}$ are less marked (Figure 4A through 4C). Depending on the fitting range and the particular sight line sample, the normalized EWs of the $C_2$ DIBs may increase slightly at large $f_{H2}$. Another striking feature of all the plots is the large dispersion in the individual sight line values of the normalized DIB EWs at a given $f_{H2}$, compared to the much smaller errors in the measurements. This cosmic scatter is discussed in Section 3.4.

In principle, this Lambda-shaped behavior could be characterized by three quantities: the slope of the rise in normalized EW for $f_{H2} < \sim 0.15$ (representing the growth rate as $f_{H2}$ increases in relatively low density gas), the location of the peak of the distribution (generally around $f_{H2} \sim 0.3$), and the slope of the decline in normalized EW for larger $f_{H2}$ (representing the destruction rate at higher densities). In practice, however, precise values for the two slopes have been difficult to determine, due to the limited numbers of sight lines in both the lowest - $f_{H2}$ and highest - $f_{H2}$ regimes and to the evident steepening of the declines (for the normal DIBs) at the highest $f_{H2}$.

The plots in Fig. 4 indicate that the slopes at small $f_{H2}$ are positive but small for all eight of the DIBs in this study. The $\lambda5797.1$ DIB and the $C_2$ DIBs exhibit the largest slopes in this regime, while several of the broader DIBs (e.g., the $\lambda6283.8$ DIB) increase very little. More dramatic differences among the DIBs are seen in the slopes at higher $f_{H2}$. While the slopes for the $C_2$ DIBs are near zero, the slopes for the 'normal' DIBs become increasingly steep, in some cases reaching values of order -3 (or less) at the highest $f_{H2}$.

The Lambda-shaped behavior seen for our larger sample of sight lines both confirms and more thoroughly defines the trends seen in several previous studies. Snow & Cohen (1974) had noted that the $\lambda5780.5$ and the $\lambda5797.1$ DIBs were weaker with respect to $E_{B-V}$ when dense clouds are present in the sight line (see also Wampler 1966, for the $\lambda4428.2$ DIB), corresponding in our case to small $W(DIB)/E_{B-V}$ at large $f_{H2}$. More observations of different regions of the sky (e.g., Strom et al. 1975; Meyer & Ulrich 1984; Adamson et al. 1991) provided further evidence. Jenniskens et al. (1994) and Sonnentrucker et al. (1997) also reported a Lambda-shaped behavior for the normalized EWs of DIBs $\lambda\lambda5780.5$, 5797.1, 6379.4, 6283.8, and 6613.6, but compared to $E_{B-V}$ rather than $f_{H2}$. A similar Lambda-shaped behavior for the normalized EWs of some DIBs ($\lambda\lambda5780.5$, 5797.1, and 6353.5) versus the fraction of atomic hydrogen (1 - $f_{H2}$) was also reported in Cami et al. (1997) for 13 nearby sight lines appearing to have only a single interstellar cloud component (most of which are included here). All previous studies that discussed the Lambda-shaped DIB distribution clearly established the sharp rise of the DIB strength with either reddening or $f_{H2}$ for $f_{H2} \leq \sim 0.3$. The systematic decrease of the DIB strength with increasing reddening (or $f_{H2}$) suggested by those previous studies is seen more clearly in the present work. Our larger data sample clearly establishes the Lambda-shaped behavior for those DIBs previously mentioned and also for the normal DIB $\lambda6196.0$. The steep drop at large $f_{H2}$ does not apply for the three $C_2$ DIBs in this study, however.

3.3 Sigma and Zeta Sight Lines

Although the EWs of DIBs $\lambda\lambda5780.5$ and 5797.1 correlate moderately well (Figure 5B), there is considerable scatter (larger than the measurement uncertainties) throughout the plot and the chi-square is large. As described previously, these two DIBs have a similar Lambda-shaped behavior with $f_{H2}$, but with different slopes at both low and high $f_{H2}$. The sigma-zeta effect (Sneden et al. 1991; Krełowski et al. 1992) refers to the significant differences observed for the $W_\lambda(5780)/W_\lambda(5797)$ ratio (or the ratio of central depths of these two DIBs in some cases), which can range from about 1 to 10 in different sight lines (e.g. Vos et al. 2011; Kos & Zwitter 2013; Welty et al. 2014). Sigma sight lines were originally defined as those where the λ5780.5 DIB is significantly deeper than the λ5797.1 DIB (as toward σ Sco), and vice versa for zeta sight lines (as toward ζ Oph). While there are also differences in far-UV extinction in those two sight lines (flat toward σ Sco, steeper than average toward ζ Oph), the physical relationship between the DIB strengths and the extinction is not understood.

The variation of the $W_\lambda(5780)/W_\lambda(5797)$ ratio as a function of $f_{H2}$ is shown in Figure 5A, with the same symbols for significant stars as used in Figure 4. As $f_{H2}$ increases from 0 to about 0.2, there is an obvious decline in the average $W_\lambda(5780)/W_\lambda(5797)$ ratio. For $f_{H2} > 0.2$, the decline becomes shallower. A similar plot is given in Weselak et al. (2004), where the strengths of the DIBs are represented by central depths (CDs) rather than EWs. In their plot, the CD_5797/CD_5780 ratio undergoes only a slight increase for $f_{H2}$ between 0.2 and 0.5, consistent with our finding. In Figure 5A, there are three noticeable outliers, relative to the well-defined decline of the $W_\lambda(5780)/W_\lambda(5797)$ ratio with increasing $f_{H2}$: HD 37903 (B1.5V, $E_{B-V}$ = 0.35 mag., $f_{H2}$ = 0.53, highlighted as △), HD 73882 (O8V, $E_{B-V}$ = 0.70 mag., $f_{H2}$ = 0.67, highlighted as ▢), and HD200775 (B2Ve, $E_{B-V}$ = 0.63 mag., $f_{H2}$ = 0.83, highlighted as ✳). This last star is the only one in the plots for which $f_{H2} > 0.8$ (though that value is based on the use of surrogates). More detailed discussion concerning these three outliers will be given in Section 4.2.2.

3.4 Cosmic Scatter

Cosmic scatter is found in all of our $f_{H2}$ plots. The dispersion among points at any given $f_{H2}$ is always much larger than the typical uncertainties for the individual measurements. The DIB behavior in the previous section thus refers to the general trends in the average normalized EWs of the DIBs, i.e. the orange squares and orange lines in the insets of each panel of Figure 4. Given the high quality of our DIB spectra and the resulting high accuracy of our measurements (Figure 2), this cosmic scatter may reflect different combinations of physical conditions in the various sight lines (and in the responses of the DIBs to those conditions). There is some evidence that the DIBs may be even more sensitive than some of the known atomic and molecular species to changes in the local physical conditions, which can occur over fairly short length scales (Cordiner et al. 2013). Moreover, high-resolution spectra of various atomic and molecular species have revealed complex velocity structure in most sight lines, with significant differences in the properties of the individual velocity components (e.g., Welty et al. 1994, 1996, 2014; Crane et al. 1995). But given the width of the DIBs, it is generally very difficult to actually tie any DIB peculiarity to a single velocity component exhibiting some other peculiar interstellar property. Unfortunately, the individual components cannot be discerned and separated for H, H$_2$, or the extinction either, and we are condemned to measure those quantities integrated over the sight line. Sight lines with similar $f_{H2}$ and $E_{B-V}$ can thus contain very different ISM environments on the scale of individual clouds. While the most extreme variations will be somewhat obscured in the sight line averages, some scatter will remain, due to differences in the relative amounts of different kinds of gas in each sight line. This discussion emphasizes the importance of detailed studies of simple sight lines, though identifying such sight lines is extremely difficult with current spectrographs. An alternative is to focus on sight lines that have some outstanding characteristic -- so that one cloud may be dominant with regard to DIB behavior -- as is the case for Herschel 36 (Dahlstrom et al 2013; Oka et al. 2013), HD 37903 (this work) or HD 62542 (Snow et al. 2002; Welty et al. in preparation).

3.5 Correlations

Pair-wise comparisons between different quantities can identify both general relationships and "discrepant" points which may illuminate the behavior of those quantities under unusual environmental conditions. Tables 2 and 3 present the Pearson correlation coefficients ($r_{\rm lin}$ for linear units and $r_{\rm log}$ for logarithmic units, respectively) and the corresponding slopes of the best-fit general trends for pair-wise correlations between $N$(H), $N$(H$_2$), $E_{B-V}$, and the eight DIBs measured in this study. While a good correlation between two quantities may suggest that they are related, the slopes can also reveal physical, chemical, or spatial relationships between the various quantities (Sec. 3 in this work, see also Welty & Hobbs 2001; Sonnentrucker et al. 2007; Sheffer et al. 2008; Welty 2014). No values of $N$(H$_2$) or $N$(H) based on surrogates were used in the correlation analyses, and, for comparisons involving $N$(H$_2$), we only considered sight lines with log[$N$(H$_2$)] > 18.5, corresponding to fully shielded H$_2$ (Savage et al. 1977). The EWs of DIBs affected by stellar line blending were excluded.

We explored two methods for fitting linear trends to the various pair-wise comparisons. The first method employed a Monte Carlo (MC) procedure, in which the raw data were modified randomly based on the measured uncertainties (Table 1). The slopes, intercepts, and correlation coefficients ($r$-values) then were determined using the simple linear fitting program LINFIT[10], with equal weighting of the data and no rejection of discrepant points. For each correlation, we obtained the average values of $r$ and of the slopes from 1,000 Monte Carlo runs, and adopted the standard deviations as errors. The second method employed an iterative fit in which the data are weighted based on the uncertainties in both quantities – using either the IDL FITEXY procedure or a similar procedure (REGRWT)[11] which we have modified to allow for equal weights and/or removal of discrepant points (> 2.5-$\sigma$). While the derived $r$-values and slopes depend to some degree on both the data sample (sight lines included, distributions in values) and the details of the fitting method (e.g., the weighting method and the treatment of 'discrepant' points), fairly similar results were often obtained from the two methods for the roughly linear relationships found for both the normal DIB - normal DIB correlations and the C$_2$ DIB - C$_2$ DIB correlations (all of which have $r \geq 0.85$). As judged via the weighted rms distances of the data from the fitted lines, the iterative procedures that consider the uncertainties in both quantities (FITEXY, REGRWT) generally yielded better fits to the relationships in log units (Table 3), while the MC/LINFIT procedure generally gave better fits to the relationships in linear units (Table 2). For a number of the comparisons with smaller $r$-values, FITEXY did not produce good fits to the main trends in the data. In such cases, allowing equal weighting of the data and/or removal of the >2.5-$\sigma$ deviants (typically 1-4 sight lines) via REGRWT generally yielded better fits; the slopes from those fits are given in square braces in Table 3.

Because of the wide diversity of characteristics of the sight lines in our sample -- e.g., with both very low and very high reddening and/or $f_{\rm H2}$ -- and because of the differences in behavior seen for the normal and the C$_2$ DIBs, however, some of the other relationships (e.g., normal DIB vs. C$_2$ DIB, or DIB vs. $E_{B-V}$, $N$(H), and/or $N$(H$_2$)) exhibit significant scatter and/or a degree of bi-modality. Those more complex relationships may reflect (for example) the differences in behavior of different DIBs in sight lines that are comparably reddened but are dominated either by diffuse atomic gas or by denser, more molecular gas. Such relationships may not be well fitted by simple linear functions y = a + b*x of a single variable x (in either linear or log units). Multivariate analyses (e.g., Lan et al. 2015; Ensor et al. 2017) will be required to disentangle the complex dependencies in such cases.

---

[10] Available within the Interactive Data Language (IDL) environment.
[11] Based on the subroutine regrwt.f, which is currently available at http://www.classification-society.org/csna/mda-sw/leiv1.f (see, e.g., Welty et al. 2003).

For the normal DIBs, our fits compare well with those obtained by Friedman et al. (2011). For the five normal DIBs in common, our average correlation coefficient with $N(H)$ is $r_{\log} = 0.79$, compared to their 0.81; both numbers are based on all suitable sight lines (no blends with other DIBs or with stellar lines in the background stars). Pruning the samples to eliminate outliers, Friedman et al. found an average correlation coefficient of 0.88. Comparing this to the small formal errors in the $r$-values (Tables 2 and 3), it appears that the main uncertainty in a given value for $r$ is not due to the measurement errors for the data, but to the selection of stars used in a given fit (and to the cosmic scatter noted above).

A number of previously examined DIB correlations are confirmed with our somewhat larger sample of stars. DIBs $\lambda\lambda 6196.0$ and $6613.6$, for example, are again found to be nearly perfectly correlated ($r_{\lin} = 0.98$) as in several previous studies (e.g. Cami et al. 1997; Moutou et al. 1999; Galazutdinov et al. 2002; McCall et al. 2010). Because the amounts of all interstellar constituents generally increase with distance, they are all positively correlated to some degree. Friedman et al. (2011) suggested using $r = 0.86$ as a threshold to indicate that two ISM quantities could be physically related. By this criterion, all the normal DIBs in this work could be related, as the mutual correlation coefficients, $r_{\lin}$, are all greater than 0.9.

Good correlations are also found among the three $C_2$ DIBs (Figure 6A, B, and C). The correlation coefficients, $r_{\lin}$, range between 0.96 and 0.91, comparable to the values found for pairs of normal DIBs. The three $C_2$ DIBs generally do not correlate as well with the normal DIBs, however[12]. The average value of $r_{lin}$ from Table 2 between $C_2$ DIBs and normal DIBs (15 pairs) is 0.77. The analogous value from Table 3 is $r_{log} = 0.74$. These results indicate that the $C_2$ DIBs represent a distinct class of DIBs well separated from the normal DIBs. Star-to-star differences in the relative strength of the redward wing of the $\lambda 4726.8$ DIB suggest that the total feature might not be a pure $C_2$ DIB, which would tend to degrade its correlation with the other pure $C_2$ DIBs. However, the good correlations observed among the three $C_2$ DIBs considered here suggest that any such degradation is minor.

Another significant difference between the $C_2$ DIBs and the normal DIBs is seen in their correlations with $N(H)$ and $N(H_2)$. Tables 2 and 3 suggest that the $C_2$ DIBs are better correlated with $N(H_2)$ while the normal DIBs have better correlations with $N(H)$. The average values of $r_{lin}$ (and $r_{log}$) between the $C_2$ DIBs and $N(H_2)$ are $0.59 \pm 0.04$ ($0.70 \pm 0.04$) while the average values between the $C_2$ DIBs and $N(H)$ are $0.44 \pm 0.06$ ($0.56 \pm 0.07$). In contrast, $r_{lin}$ (and $r_{log}$) between the normal DIBs and $N(H_2)$ are $0.45 \pm 0.02$ ($0.60 \pm 0.08$) while the average values for the normal DIBs and $N(H)$ are $0.62 \pm 0.04$ ($0.79 \pm 0.04$).

The best correlation between a DIB and $N(H)$ is for the $\lambda 5780.5$ DIB ($r_{\log} = 0.86$), which is the only correlation that reaches the $r = 0.86$ threshold[13]. On the other hand, the best correlation of a $C_2$ DIB or a normal DIB with $N(H_2)$ is for $\lambda 4963.9$, where $r_{log} = 0.74$, so no DIB exhibits a pure association with $N(H_2)$. While both the normal and the $C_2$ DIBs are moderately well correlated with $E_{B-V}$, ($r_{lin} = 0.76 - 0.89$), the DIBs $\lambda\lambda 4726.8, 4963.9, 5780.5, 5797.1, 6196.0$, and $6613.6$ reach the level $r > 0.85$, with the first two being the most closely related to the dust ($r_{lin} = 0.89$ and 0.88, respectively).

The slopes of the pair-wise correlations are also reported in Tables 2 and 3, where the abscissa in the correlation is the entity on the right and the ordinate is the entity to the left (e.g. the slope for $W_\lambda(6283)$ as the abscissa and $W_\lambda(6196)$ as the ordinate is 0.05 for the linear

---

[12]Besides the five "normal DIBs" beyond 5000Å considered in this paper, we include in this category $\lambda\lambda 5487.7$, 5705.1, and 6204.5 from Friedman et al. (2011) (see Section 2.3.3 and 3.2). We note the $\lambda 5797.1$ DIB has a better correlation with the $C_2$ DIB $\lambda 4726.8$. Please refer to the discussion in Section 4.4.

[13]When the outliers are left out in the $N(H)$ verses $\lambda 5780.5$ relationship, the correlation is much better (Figure 3A gives r = 0.972). The correlation in Figure 3A is the correlation to use when the $\lambda 5780.5$ DIB is to be used as a surrogate for $N(H)$. Care must be taken not to use this surrogate when there is indication of a strong radiation field in the sight line. Please refer to the discussion in Section 4.4

correlation (Table 2)). The slopes for log-log correlations that do not involve $N(H_2)$ are generally of order unity, consistent with roughly linear relationships. On the other hand, the slopes versus $N(H_2)$ are always smaller than 0.6, possibly due to the very effective accumulation of $H_2$ after self-shielding is established.

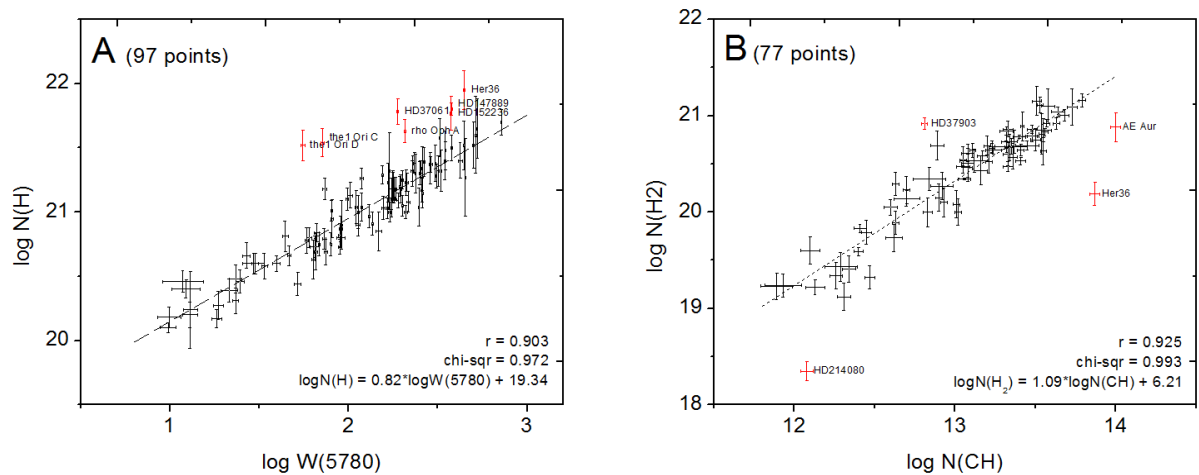

Figure 3. Correlations for log $W_\lambda(5780)$ vs. log $N(H)$ (left) and log $N(CH)$ vs. log $N(H_2)$ (right). Outliers, marked as red points, are excluded from the best fit. The fits were done as described in Section 3.5, and outliers in red were excluded. Both relationships have high correlation coefficients and the chi-square values indicate that the scatter can be explained by the measurement errors, while the outliers are likely due to special environmental conditions in the sight lines.

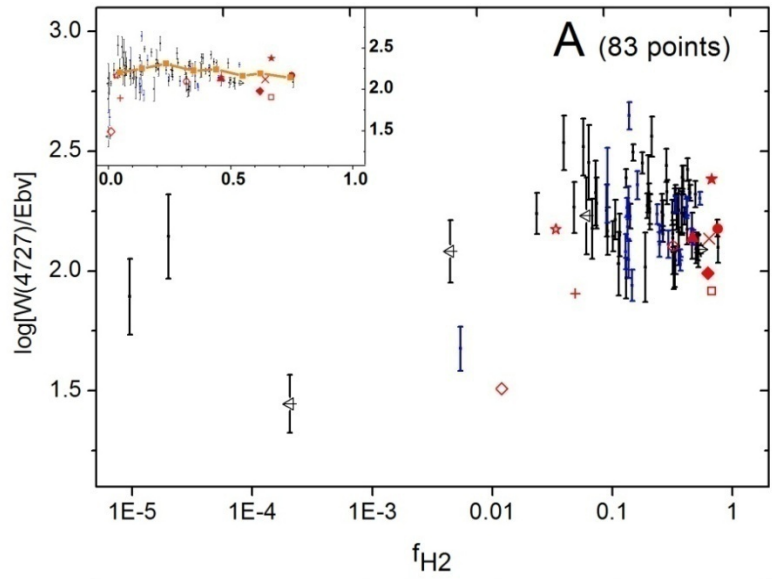
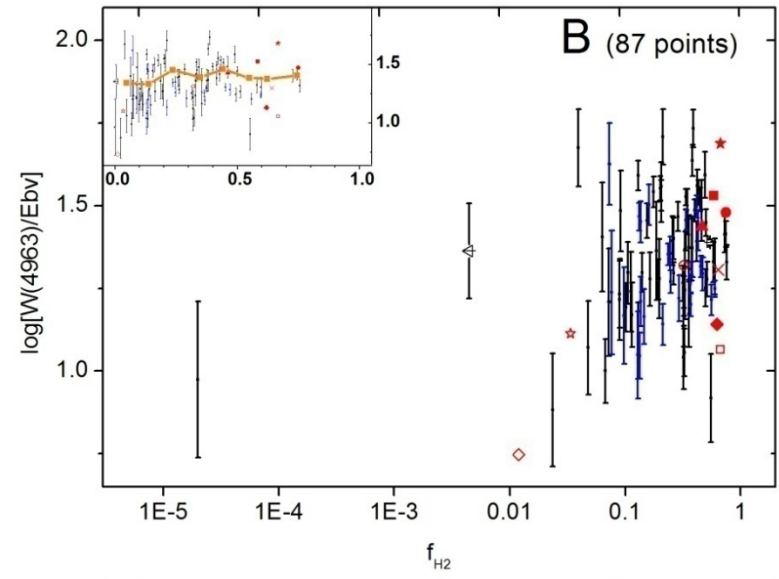
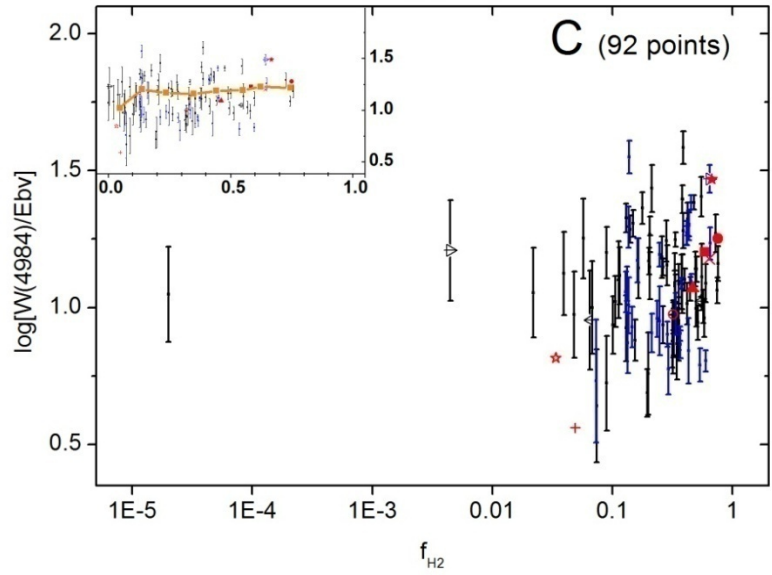
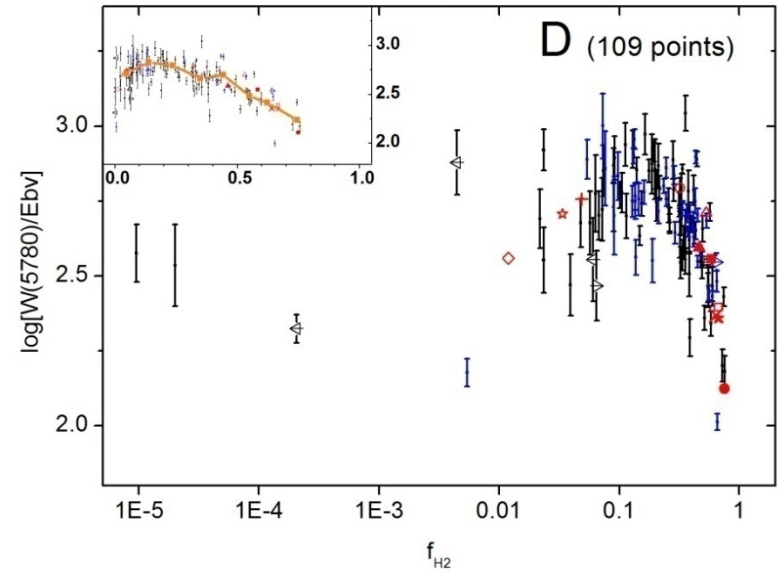

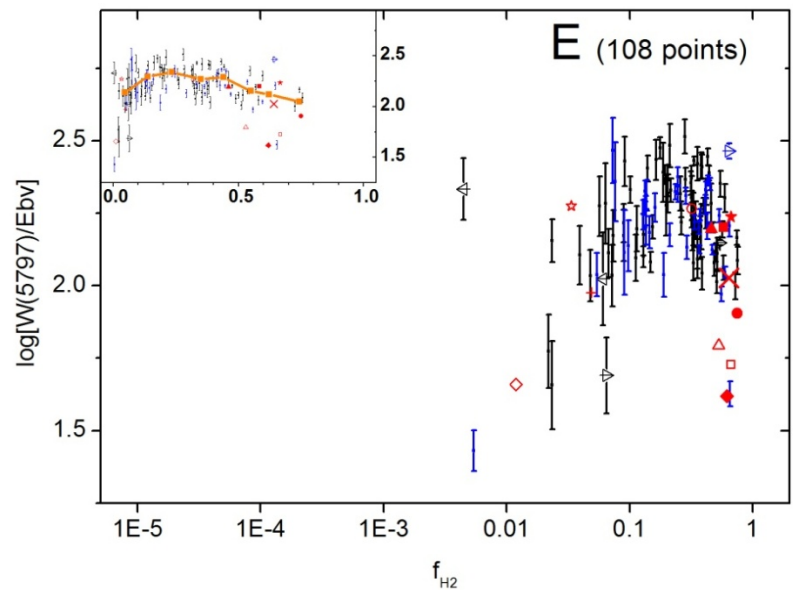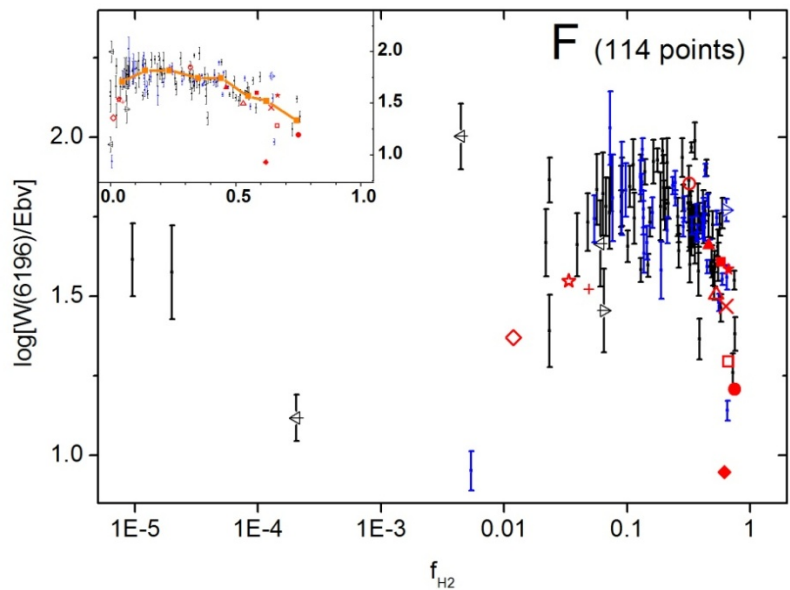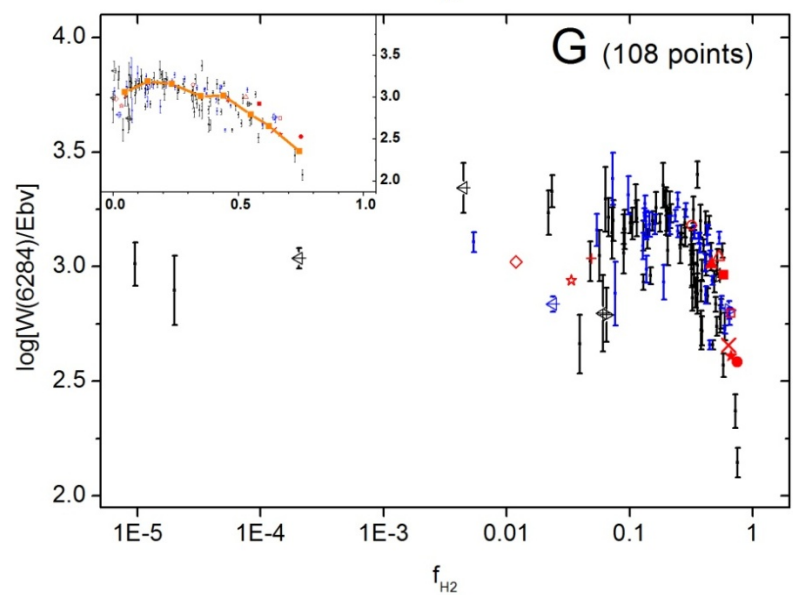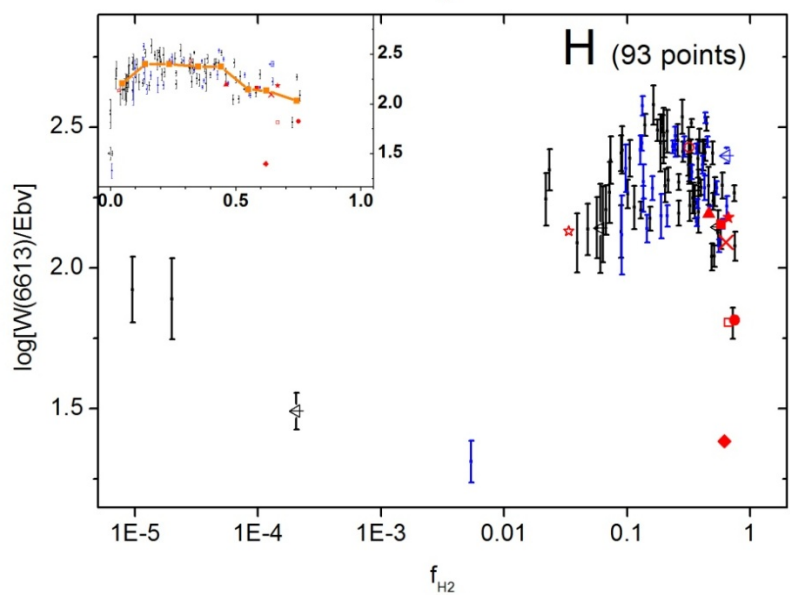

Figure 4. Behavior of the normalized EWs of 8 DIBs as a function of $f_{H2}$ in log-log units. The $f_{H2}$ values of the black points are directly calculated from $N(H)$ and $N(H_2)$, whereas the blue points are based on the use of surrogates of $W_\lambda(5780)$ for $N(H)$ and $N(CH)$ for $N(H_2)$. Arrows pointing toward left and right represent upper and lower limits for the $f_{H2}$ values. The small inset in the upper left corner of each plot represents the same data with the abscissa plotted on a linear scale. Inside these insets, the orange squares represent the averages in each of the eight bins of a constant width of $f_{H2} = 0.1$, and the orange line connecting the orange squares shows the general trend of the average of the data points. Special stars are noted as follows: HD 147165 (Sig-Sco, ✚), HD 149757 (Zet-Oph, **X**), NGC2024-1 (●), HD 204827 (★), BD -14 5037 (■), Cyg OB2 #5(▲), HD 37903 (△), HD 73882 (□), HD 37061 (◇), HD 62542 (◆), Herschel 36 (☆), HD 183143 (○), and HD 200775 (✻), as explained in Section 3.2.

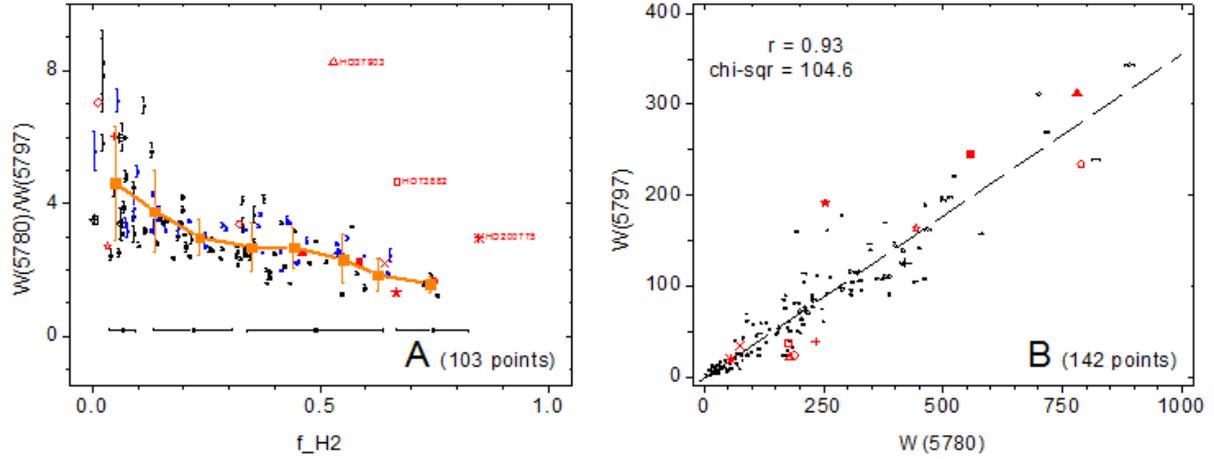

Figure 5. A: $W_\lambda(5780)/W_\lambda(5797)$ ratio as a function of $f_{H2}$; B: correlation between DIBs $\lambda\lambda 5780.5$ and 5797.1. We follow the same symbol-system used in Figure 4. Typical errors for different ranges of $f_{H2}$ are shown at the bottom of panel A. We note that there is only one measurement at $f_{H2} > 0.8$, HD 200775 (✳), in front of the reflection nebula NGC 7023 (Sellgren et al. 1983), whose $f_{H2}$ value is based on the use of surrogates. The largest $W_\lambda(5780)/W_\lambda(5797)$ ratio is observed in the sight line of HD 141637 at $f_{H2} = 0.02$, and the smallest ratio is observed in the sight line of HD 24534 at $f_{H2} = 0.76$. Orange squares represent the averages of each bin of width of $f_{H2} = 0.1$ as in the insets of Figure 4. We also show the standard deviation of the data points in each bin, which is significantly larger than the typical uncertainty in each measurement.

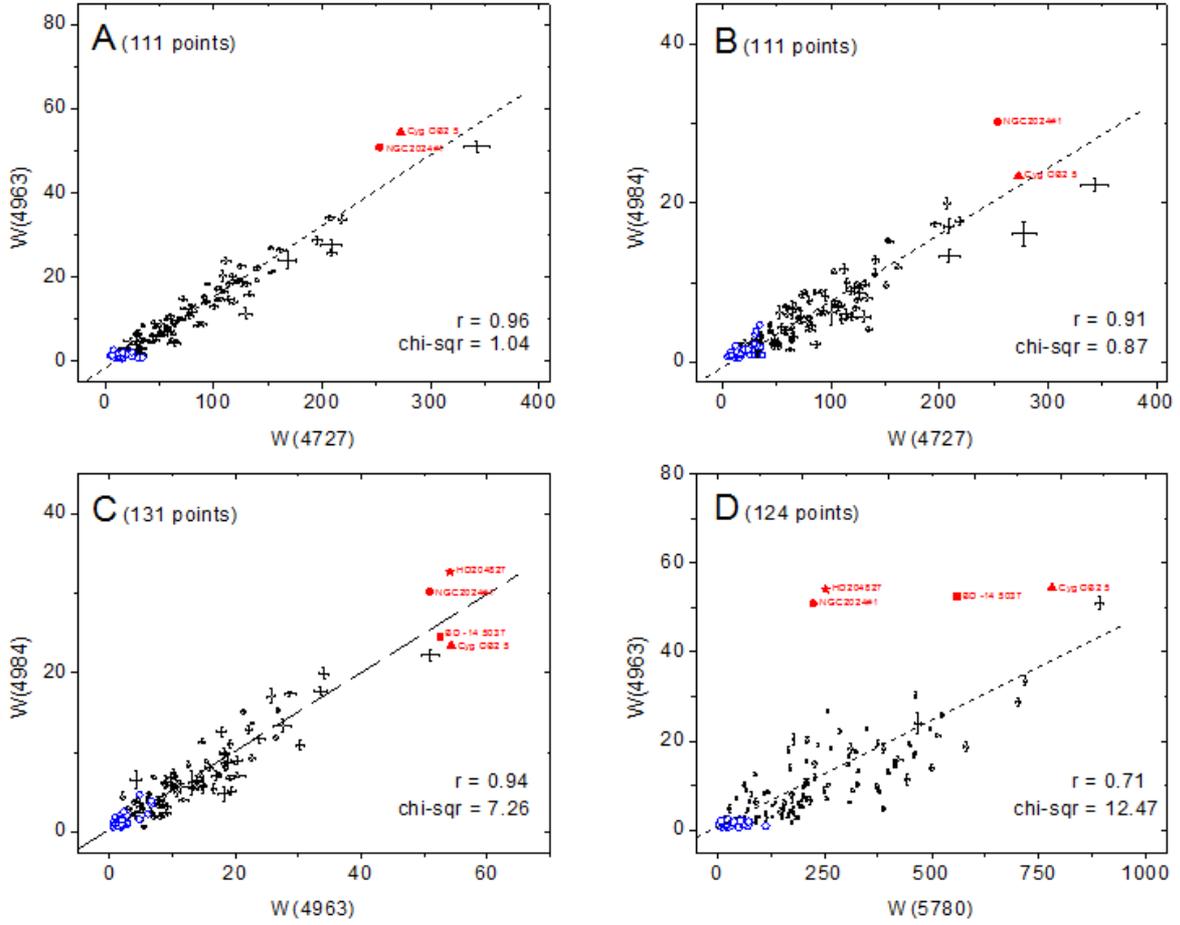

Figure 6. Correlations among C$_2$ DIBs (Panel A, B, and C) and between the normal DIB $\lambda$5780.5 and the C$_2$ DIB $\lambda$4963.9 (Panel D). The fits were derived as described in Section 3.5. Blue circles indicate upper limits in either quantity. The four sight lines with very strong C$_2$ DIBs (NGC2024-1, HD 204827, BD -14 5037, and Cyg OB2 #5) are highlighted. We also give chi-squares for each fit, as a comparison between residuals and measurement uncertainties.

Table 2: Correlation Coefficients and Slopes for Pair-Wise Correlations in Linear Units[a,b]

| | $W_\lambda(4726)$ | $W_\lambda(4963)$ | $W_\lambda(4984)$ | $W_\lambda(5780)$ | $W_\lambda(5797)$ | $W_\lambda(6196)$ | $W_\lambda(6283)$ | $W_\lambda(6613)$ | $E_{B-V}$ | $N(H)$[c] | $N(H_2)$[c,d] |
|---|---|---|---|---|---|---|---|---|---|---|---|
| $W_\lambda(4726)$ | | 5.54±0.09 | 9.75±0.23 | 0.30±0.01 | 0.83±0.01 | 2.58±0.03 | 0.11±0.01 | 0.64±0.01 | 150.4±2.6 | 1.59±0.34 | 6.81±1.08 |
| $W_\lambda(4963)$ | 0.96±0.01 | | 1.81±0.04 | 0.05±0.01 | 0.14±0.01 | 0.42±0.01 | 0.02±0.01 | 0.10±0.01 | 22.6±0.4 | 0.20±0.05 | 1.25±0.17 |
| $W_\lambda(4984)$ | 0.91±0.01 | 0.94±0.01 | | 0.05±0.01 | 0.07±0.01 | 0.17±0.01 | 0.01±0.01 | 0.01±0.01 | 11.7±0.3 | 0.12±0.03 | 0.55±0.11 |
| $W_\lambda(5780)$ | 0.81±0.01 | 0.71±0.01 | 0.58±0.01 | | 2.40±0.01 | 8.71±0.01 | 0.40±0.01 | 2.16±0.01 | 377.0±4.0 | 5.49±0.93 | 13.7±2.6 |
| $W_\lambda(5797)$ | 0.90±0.01 | 0.85±0.01 | 0.74±0.01 | 0.93±0.01 | | 3.22±0.01 | 0.14±0.01 | 0.80±0.01 | 143.1±1.8 | 1.95±0.34 | 6.15±0.13 |
| $W_\lambda(6196)$ | 0.83±0.01 | 0.76±0.01 | 0.61±0.01 | 0.97±0.01 | 0.97±0.01 | | 0.05±0.01 | 0.25±0.01 | 42.2±0.6 | 0.58±0.10 | 1.70±0.32 |
| $W_\lambda(6283)$ | 0.75±0.01 | 0.69±0.01 | 0.47±0.01 | 0.96±0.01 | 0.90±0.01 | 0.94±0.01 | | 4.90±0.02 | 836.0±12.1 | 10.4±1.9 | 35.1±5.9 |
| $W_\lambda(6613)$ | 0.83±0.01 | 0.72±0.01 | 0.60±0.01 | 0.96±0.01 | 0.96±0.01 | 0.98±0.01 | 0.92±0.01 | | 160.9±2.4 | 2.14±0.54 | 8.32±1.36 |
| $E_{B-V}$ | 0.89±0.01 | 0.88±0.01 | 0.76±0.01 | 0.83±0.01 | 0.88±0.01 | 0.86±0.01 | 0.81±0.01 | 0.83±0.01 | | 0.01±0.01 | 0.04±0.01 |
| $N(H)$[c] | 0.52±0.06 | 0.40±0.06 | 0.39±0.06 | 0.69±0.05 | 0.62±0.05 | 0.58±0.07 | 0.62±0.06 | 0.57±0.07 | 0.64±0.05 | | 1.01±0.39 |
| $N(H_2)$[c,d] | 0.57±0.05 | 0.66±0.05 | 0.53±0.07 | 0.39±0.07 | 0.46±0.07 | 0.45±0.08 | 0.41±0.06 | 0.53±0.07 | 0.71±0.04 | 0.22±0.08 | |

[a] Upper right section for slopes and lower left section for correlation coefficients.
[b] Using MC/LINFIT procedure, with no outliers omitted from the fitting.
[c] In the unit of $10^{20}$ cm$^{-2}$.
[d] Exclude sight lines with $\log[N(H_2)] < 18.5$.

Table 3: Correlation Coefficients and Slopes for Pair-Wise Correlations in Logarithmic Units[a,b]

| | $W_\lambda(4726)$ | $W_\lambda(4963)$ | $W_\lambda(4984)$ | $W_\lambda(5780)$ | $W_\lambda(5797)$ | $W_\lambda(6196)$ | $W_\lambda(6283)$ | $W_\lambda(6613)$ | $E_{B-V}$ | $N(H)$ | $N(H_2)$[c] |
|---|---|---|---|---|---|---|---|---|---|---|---|
| $W_\lambda(4726)$ | | 0.84±0.01 | 0.83±0.01 | 0.95±0.01 | 0.93±0.01 | 0.83±0.01 | 0.90±0.01 | 0.80±0.01 | [1.06±0.04] | [1.00±0.04] | [0.35±0.04] |
| $W_\lambda(4963)$ | 0.93±0.01 | | 1.04±0.01 | [1.1±0.1] | 1.01±0.01 | [1.10±0.07] | [0.99±0.08] | [1.00±0.07] | 1.18±0.01 | [1.4±0.1] | 0.57±0.01 |
| $W_\lambda(4984)$ | 0.88±0.01 | 0.86±0.01 | | [1.07±0.09] | [1.04±0.05] | [1.0±0.1] | [1.02±0.09] | [0.98±0.07] | 1.22±0.02 | [1.0±0.2] | 0.57±0.04 |
| $W_\lambda(5780)$ | 0.81±0.01 | 0.73±0.01 | 0.63±0.01 | | 0.85±0.01 | 0.98±0.01 | 0.98±0.01 | 0.93±0.01 | [1.03±0.05] | 1.16±0.03 | 0.31±0.01 |
| $W_\lambda(5797)$ | 0.91±0.01 | 0.86±0.01 | 0.78±0.01 | 0.93±0.01 | | 1.05±0.01 | 1.11±0.01 | 0.94±0.01 | 1.19±0.01 | [1.18±0.06] | [0.43±0.04] |
| $W_\lambda(6196)$ | 0.85±0.01 | 0.79±0.01 | 0.67±0.01 | 0.95±0.01 | 0.96±0.01 | | 0.99±0.01 | 0.96±0.01 | [1.09±0.04] | [1.01±0.05] | [0.33±0.04] |
| $W_\lambda(6283)$ | 0.71±0.01 | 0.65±0.01 | 0.50±0.01 | 0.93±0.01 | 0.85±0.01 | 0.88±0.01 | | 0.88±0.01 | [1.07±0.05] | [1.03±0.06] | 0.37±0.04 |
| $W_\lambda(6613)$ | 0.87±0.01 | 0.79±0.01 | 0.70±0.02 | 0.97±0.01 | 0.97±0.01 | 0.97±0.01 | 0.87±0.01 | | 1.24±0.01 | [1.37±0.07] | 0.40±0.01 |
| $E_{B-V}$ | 0.82±0.02 | 0.85±0.02 | 0.78±0.02 | 0.86±0.02 | 0.87±0.01 | 0.86±0.02 | 0.82±0.02 | 0.87±0.01 | | 1.11±0.04 | [0.36±0.04] |
| $N(H)$ | 0.64±0.03 | 0.55±0.04 | 0.48±0.04 | 0.86±0.02 | 0.77±0.02 | 0.76±0.07 | 0.80±0.02 | 0.77±0.02 | 0.79±0.02 | | [0.22±0.04] |
| $N(H_2)$[c] | 0.68±0.02 | 0.74±0.02 | 0.70±0.03 | 0.60±0.02 | 0.71±0.02 | 0.61±0.02 | 0.46±0.02 | 0.64±0.02 | 0.82±0.03 | 0.42±0.04 | |

[a] Upper right section for slopes and lower left section for correlation coefficients.
[b] Using FITEXY and/or REGRWT procedures; cases where outliers were omitted from the fitting are in square braces; less certain values are given to one decimal place
[c] Exclude sight lines with $\log[N(H_2)] < 18.5$.

## 4. DISCUSSION

### 4.1 Diffuse Atomic and Molecular Gas

Snow & McCall (2006) proposed a distinction between diffuse atomic, diffuse molecular, and translucent interstellar material based not on $E_{B-V}$ or $A_V$, but on the relative abundances of H and $H_2$, and the dominant repositories of carbon. In diffuse atomic gas, hydrogen is primarily in neutral atomic form, carbon is primarily singly ionized, and the abundances of even simple diatomic molecules are quite low. In diffuse molecular gas, most of the hydrogen has been converted to $H_2$, but most of the carbon is still in $C^+$ (with small but increasing fractions in C, CO, and other simple molecules). In translucent material, C and/or CO become the dominant carbon species. Photoprocesses such as photoionization and photodissociation are significant in both diffuse atomic and diffuse molecular gas, though less so in the latter, due to increased shielding of the ambient UV radiation by dust and, for CO and $H_2$, by self-shielding via saturated UV absorption lines (Glassgold et al. 1987; Draine & Bertoldi 1996).

Models have suggested that the transition from H to $H_2$ that occurs when self-shielding becomes effective happens fairly rapidly; measurements of $N$(H) and $N(H_2)$ in the Galactic ISM indicate that the molecular fraction $f_{H2}$ rises dramatically at $E_{B-V} \sim 0.08$ (Savage et al. 1977). In principle, $f_{H2}$ thus can be used to distinguish between diffuse atomic and diffuse molecular gas. Large $f_{H2}$ will be found when the sight line is dominated by the cores of diffuse molecular clouds, but the actual $f_{H2}$ cannot be determined for individual velocity components because the strong low-$J$ lines of $H_2$ are generally broader than the typical velocity separations between those components (Morton 1975). Based on higher resolution spectra of $H_2$ in low-$N(H_2)$ sight lines obtained with *IMAPS* (Jenkins et al. 2000), on even higher resolution spectra of other atomic and molecular species that are well correlated with $H_2$ (e.g., Welty & Hobbs 2001), and on the $b$-values inferred from higher-$J$ $H_2$ lines (e.g., Jensen et al. 2010), it is likely that multiple components are present for most of the sight lines in which $H_2$ is detected. The highest integrated values of $f_{H2}$ in the *FUSE* survey of highly reddened sight lines are between 0.65 and 0.76 (Rachford et al. 2002, 2009), slightly higher than the highest values found for the typically less reddened sight lines observed with *Copernicus* (Savage et al. 1977).

Clues to the behavior and properties of DIBs in these interstellar environments may be obtained from comparisons with the corresponding behavior of known atomic and molecular species. While the roughly quadratic relationships for the trace neutral species Na and K versus $H_{tot}$ are reasonably consistent with expectations from considerations of ionization balance (including both radiative and grain-assisted recombination), the shallower dependencies for Fe and Ca reflect the increasing depletions of iron and calcium in denser gas (Welty & Hobbs 2001; Welty et al. 2003). Roughly linear relationships with $H_{tot}$ are found for the dominant ions of little-depleted elements (e.g., O, $P^+$, $Zn^+$), while somewhat shallower trends are seen for the dominant ions of more severely depleted elements (e.g., $Mg^+$, $Ti^+$, $Fe^+$), again reflecting the increasing depletions of those elements with increasing density (and thus increasing $f_{H2}$; e.g., Cardelli 1994; Jensen & Snow 2007a, 2007b; Welty & Crowther 2010).

Because different molecules occupy different locations in the chemical reaction networks and have different photodissociation thresholds, differences in spatial distribution are to be expected (e.g., Federman et al. 1994; Pan et al. 2005). Those differences in distribution can be reflected in the slopes of correlations between those molecules and $H_2$, with steeper positive slopes for species that are more centrally concentrated in denser parts of the clouds (e.g., Sheffer et al. 2008; Welty et al., in preparation). For example, CH, a "first-generation" molecule whose production (in the standard picture of gas-phase chemistry) is initiated by a reaction between $C^+$ and $H_2$, is nearly linearly related to $H_2$ (e.g. Danks et al. 1983), and is relatively broadly distributed (Pan et al. 2005). On the other hand, "second-" and "third-generation" molecules (e.g., $C_2$, $C_3$, CN, and CO), which require precursor molecules containing heavy atoms for their production, and are thus generally more centrally concentrated in denser regions, exhibit somewhat steeper relationships with $H_2$ (Sonnentrucker et al. 2007; Sheffer et al. 2008).

There have been a number of indications that the DIBs also depend on local environmental conditions, with some differences in response for individual DIBs. Herbig (1995) found a roughly linear relationship between $W_\lambda(5780)$ and $N(H)$, with no residual dependence on $N(H_2)$, consistent with that DIB tracing primarily atomic gas and with its carrier being a dominant species (ionization state) in such gas. Friedman et al. (2011) and Lan et al. (2015) found similar relationships with $N(H)$ for a number of additional DIBs as well (though with slightly different slopes and degrees of correlation); Lan et al. also found slight residual correlations or anti-correlations with $N(H_2)$ for a number of the DIBs in their sample. By analogy with the relationships found among the known molecular species, the differences in slopes for the correlations between the EWs of various DIBs and the column densities of atomic hydrogen and of other known species (in log-log plots) may imply differences in the spatial distributions of the DIB carriers, with a possible connection with the widths of the individual DIBs (Welty 2014; Lan et al. 2015). Correlations between the $W_\lambda(5780)/W_\lambda(5797)$ ratio and both the $N(H)/E_{B-V}$ ratio and $f_{H2}$ suggest that the $\lambda 5797.1$ DIB traces somewhat denser regions than the $\lambda 5780.5$ DIB (Krelowski et al. 1999; Weselak et al. 2004, 2008). While the enhancement of the $C_2$ DIBs, with respect to the $\lambda 6196.0$ DIB, in sight lines where the $C_2$ molecule is more abundant (Thorburn et al. 2003) is at least partially due to the decline of the normalized EW of the $\lambda 6196.0$ DIB in such sight lines (see Figure 4F), our observation that the normalized EWs of the $C_2$ DIBs do not decline at large $f_{H2}$ values nonetheless indicates that they may trace denser regions of the clouds than the normal DIBs. Whether this relative enhancement is due to a more efficient formation of the carriers of the $C_2$ DIBs in the denser regions is not yet clear.

The DIBs also appear to respond to differences in the local radiation field. Observations that $W(DIB)/E_{B-V}$ for some DIBs decrease with increasing $E_{B-V}$ for sight lines probing moderately dense regions provided early evidence that those DIBs are located primarily in the more diffuse outer layers of the clouds (the "skin effect"; Snow & Cohen 1974; Strom et al. 1975; Meyer & Ulrich 1984) and suggested that radiation and shielding effects are important. To varying degrees, the DIBs appear to be weaker in regions where the UV radiation fields are stronger than average (e.g., Orion Trapezium and parts of Sco-Oph; Herbig 1993; Vos et al. 2011). The relatively weak DIBs found in most of the sight lines observed so far in the Large and Small Magellanic Clouds appear to be due to the combined effects of the lower metallicities, typically somewhat enhanced radiation fields, and additional metallicity-related effects (Cox et al. 2006, 2007; Welty et al. 2006). Strong local IR radiation fields can apparently modify the profiles of some of the DIBs (e.g., toward Herschel 36; Dahlstrom et al. 2013), providing clues to the characteristics of the carrier molecules (Oka et al. 2013; Huang & Oka 2015).

4.2 Formation, Destruction, and Modification of DIB Carriers

In principle, the differences in DIB strength in different environments must reflect the various processes involved in the formation, destruction, and/or modification of the DIB carrier molecules in those environments. Comparisons of the DIB strengths with local physical conditions (e.g., UV/optical/IR radiation fields, densities, fractional ionization, molecular fraction, and metallicity) thus may provide constraints on both the relevant processes and the nature of the carriers. If the carrier molecules generally contain more than five heavy atoms (i.e., other than H; e.g., Huang & Oka 2015), then it seems unlikely that the carriers can be formed "bottom-up" via gas-phase chemistry in the relatively diffuse clouds in which most of the carriers appear to reside, given (for example) the observed trend of decreasing abundances of carbon chain molecules of increasing size in those clouds (Maier et al. 2002). It seems more likely that the carriers are formed "top-down", as fragments of larger molecules or dust grains, which are then able to survive in the diffuse clouds (Duley 2006; Zhen et al. 2014). Other processes (e.g., ionization, hydrogenation or de-hydrogenation, depletion onto grains, and other chemical reactions) may then modify the carriers or remove them from the gas phase to produce changes in the observed DIB strengths.

The radiation field, especially in the UV, is a key environmental factor in the diffuse ISM, given its capability to ionize atoms and to ionize and dissociate molecules. As noted above, the presence of a strong radiation field has been used to explain the anomalously low abundances of molecular hydrogen, trace neutral species, and some DIBs in the significantly reddened theta Orionis sight lines (Savage et al. 1977; Herbig 1993; Welty & Hobbs 2001). On the other hand, absorption of the incident UV radiation in the outer layers of a cloud, either by dust or by strongly saturated absorption lines, may shield atoms and molecules in the inner parts of the cloud; both mechanisms are significant for the conversion of H to $H_2$, for example. It has been hypothesized that the carriers of different DIBs may need different amounts of either shielding from or exposure to the ambient radiation field, which then leads to their different behavior when subjected to fields of different strength and/or shape. The nature of the shielding is unclear, however. There are no observations of saturated DIBs, or even any securely confirmed DIBs, in the UV part of the spectrum. The shielding for DIBs thus would have to be continuous shielding over the UV spectrum rather than line shielding as for $H_2$ or CO (Solomon & Wickramasinghe 1969; van Dishoeck & Black 1988; Warin et al. 1996). Sonnentrucker et al. 1997 developed a simple ionization model for the DIB carriers, based on the analysis of the Lambda-shaped distribution, which yielded estimates of DIB ionization potentials (IPs). Those IPs were found to be consistent with those of neutral or singly ionized PAH molecules, which have been proposed to be DIB carriers (and which can turn into fullerenes via sequential dehydrogenation and $C_2$ fragment losses (Zhen et al. 2014)). The fullerenes are also promising DIB carriers, especially after the identification of $C_{60}^+$ as the carrier of several DIBs (Campbell et al. 2015). The low abundance of the fullerenes means, however, that they are likely not the only kind of DIB carrier, especially for the strong bands studied here (Omont 2016). The limits on carrier size for the $\lambda$5797.1 and $\lambda$5780.5 DIBs inferred by Oka et al. (2013) and Huang & Oka (2015) also raise the question of whether PAHs and fullerenes can be the carriers for those DIBs, and more generally, what the dominant ionization state of the DIB carriers might be in any given part of the cloud. It is not certain that the radiation field is the only factor governing the behavior of the DIBs, as other parameters, such as density, may contribute as well.

4.2.1 Behavior of Normalized EWs of DIBs vs. $f_{H2}$

Using $f_{H2}$ as a tracer of the integrated interstellar environment along different sight lines (see Section 3.1), we confirm, with an extensive dataset covering a wide range of values of $f_{H2}$, the Lambda-shaped behavior for all five normal DIBs in our sample, as was noted for a few normal DIBs in previous works (Sonnentrucker et al. 1997, 1999). The normalized EWs of different normal DIBs generally increase from $f_{H2} = 0$ to ~ 0.15, decrease for $f_{H2} > ~ 0.3$ (to varying degrees), and peak at different intermediate $f_{H2}$. While the differences in both slopes and peak $f_{H2}$ suggest that the DIBs arise from different carriers, it is not yet clear whether those differences are due to differences in the formation, destruction, or modification of the carriers. The differences do suggest, however, that the various DIBs may preferentially arise in somewhat different parts of the clouds, characterized by differences in local physical conditions.

In the model of Sonnentrucker et al. (1997; see also Cami et al. 1997) the Lambda-shaped trends for the normalized EWs of four DIBs ($\lambda\lambda$5780.5, 5797.1, 6379.4, and 6613.6), versus $E_{B-V}$, are interpreted as changes in the charge states of their carriers: ionized in the more diffuse outer layers of the clouds, where the UV field is relatively strong (and where the normal DIBs arise), and neutral in the more shielded interiors (where the normal DIBs are suppressed). For their sample of four DIBs, the curves for the $\lambda$5780.5 and the $\lambda$5797.1 DIBs peak at the smallest and largest $E_{B-V}$ respectively, so they concluded that the carrier of the $\lambda$5780.5 DIB is the most resistant to UV radiation while the carrier of the λ5797.1 DIB is the least resistant to UV radiation in their data sample.

The Lambda-shaped trends for normal DIBs with increasing $f_{H2}$ in Figure 4 (in linear units, see the insets in each panel of the figure) thus appear to be qualitatively consistent with the proposed ionization model: as the postulated ionized carriers become neutral in higher density regions, the normalized EWs of the corresponding DIBs would decrease (see Sonnentrucker et al. 1997). However, a corollary of this hypothesis is that the resulting neutral molecules should produce a new set of absorption lines tracing the denser gas. Hobbs et al. (2008, 2009) show that there are some DIBs that are stronger in the molecule-rich clouds toward HD 204827 than in the more diffuse clouds toward the comparably reddened HD 183143, and the reverse is also true. It will be interesting to see if additional DIBs are found in other significantly reddened sight lines, or if any anti-correlations are found between DIBs, both of which could be consistent with changes in the ionization states of the DIB carriers. Comparisons of the behavior of the DIBs with that of known atomic and molecular species (Section 4.3 below) appear to indicate additional shortcomings with the ionization/shielding picture, however.

4.2.2 Ratio of $W_\lambda(5780)/W_\lambda(5797)$

Figure 5A shows both the observed general decline in the $W_\lambda(5780)/W_\lambda(5797)$ ratio with increasing $f_{H2}$ and several apparently "discrepant" points. The overall decline reflects the typical behavior of the two individual DIBs. In our data set, the normalized EW of the λ5797.1 DIB grows faster than the λ5780.5 DIB at $f_{H2} < 0.15$, which suggests that the carrier of the λ5797.1 DIB is more sensitive to change in the ISM environment at low $f_{H2}$. This difference produces the steep drop of the $W_\lambda(5780)/W_\lambda(5797)$ ratio for low (but increasing) $f_{H2}$ (Figure 5A). Above $f_{H2} \sim 0.3$, where the Lambda-shaped curves for both DIBs peak, the normalized EW of the λ5780.5 DIB decreases at a faster rate than that of the λ5797.1 DIB, so the $W_\lambda(5780)/W_\lambda(5797)$ ratio continues to decline (more gradually) with increasing $f_{H2}$. This also implies that the carrier of the λ5797.1 DIB is favored, compared to the λ5780.5 DIB, to survive deeper within the clouds, at higher $f_{H2}$. The largest $W_\lambda(5780)/W_\lambda(5797)$ ratios thus seem to be restricted to $f_{H2} < 0.2$ (except for the three outliers), and the smallest ratios are found only for $f_{H2} > 0.5$. Moderate $W_\lambda(5780)/W_\lambda(5797)$ ratios are found at all values of $f_{H2}$, which may be a consequence of averaging multiple clouds (affecting both DIB EWs and $f_{H2}$ in those sight lines).

In the ionization/shielding picture of DIB behavior, the sigma-zeta effect has been typically attributed to the inferred greater sensitivity of the λ5797.1 DIB to ionizing radiation (e.g., Sonnentrucker et al. 1997). In contrast to the monotonic decline of the $W_\lambda(5780)/W_\lambda(5797)$ ratio versus $f_{H2}$ found in this study, Vos et al. (2011) found a minimum in that ratio, versus $E_{B-V}$, for $E_{B-V} \sim 0.25$, which they interpreted as representing either optimal shielding for the carrier of the λ5797.1 DIB, or a point where shielding for the presumed ionized carrier of the λ5780.5 DIB is too strong. Using estimates for the local UV field derived from a simple geometric model of the significant UV sources in the Upper Scorpius region and the observed molecular abundances, Vos et al. also found that both low values of the $W_\lambda(5780)/W_\lambda(5797)$ ratio and high values of $f_{H2}$ appear to be associated with relatively weak UV fields.

Examination of some individual sight lines where the radiation field is known to differ significantly from the typical interstellar field and/or where the $W_\lambda(5780)/W_\lambda(5797)$ ratio appears to be anomalous may yield further insights into the effects of different radiation fields on the DIB carriers, as follows:

1) The star HD 37903 (B1.5V, $E_{B-V} = 0.35$ mag, $f_{H2} = 0.53$, highlighted as △), is the exciting star in front of the reflection nebula NGC 2023. The dominant cloud containing molecular hydrogen is near the star and vibrational levels of $H_2$ pumped by UV photons have been observed, indicating the cloud is bathed in a strong UV radiation field (Meyer et al. 2001; Gnaciński 2011).

This star is the most significant outlier in Fig. 5A, more than 6-σ away from the envelope of the general trend. In Figures 4E and 4F, the normalized EW of the $\lambda$5797.1 DIB is weaker toward HD 37903 by ~ 50% compared to other sight lines at similar $f_{H2}$, while the normalized EW of the $\lambda$5780.5 DIB fits the general trend reasonably well. The anomalous $W_\lambda(5780)/W_\lambda(5797)$ ratio observed in this sight line is thus due to the weaker $\lambda$5797.1 DIB (compared to $E_{B-V}$). We do not know if the weak $W_\lambda(5797)$ arises in the dominant molecular hydrogen cloud, but the rarity of both vibrationally excited H$_2$ and an anomalously weak $\lambda$5797.1 DIB makes it likely.

2) The Orion Trapezium sight lines are also known for harboring clouds with significant extinction in a strong radiation field (Krełowski & Sneden 1995), which is from the combination of an O-type star (HD 37022) and flat UV extinction (Fitzpatrick & Massa 1990). In those sight lines, the C$_2$ DIBs and the $\lambda$5797.1 DIB are absent or very weak, yielding a large $W_\lambda(5780)/W_\lambda(5797)$ ratio (Table 1). However, the absence of molecular hydrogen (e.g. log $N$(H$_2$) < 17.65 for HD 37022, Savage et al. 1977) and the corresponding upper limits for $f_{H2}$ ($f_{H2}$ < 0.01; see Table 1) makes these sight lines consistent with the trend in Figure 5A. The strength of the $\lambda$5780.5 DIB is also weakened compared to other sight lines with similar reddening, but to a lesser extent than $W_\lambda(5797)$, while $W_\lambda(6283)$ seems relatively unaffected by the strong UV radiation field. The sight line toward HD 37061 (near the Trapezium, B1V, $E_{B-V}$ = 0.52 mag, $f_{H2}$ = 0.01, highlighted as ◇) also exhibits weak absorption from trace neutral species and from vibrationally excited H$_2$, suggestive of an enhanced UV field there as well.

3) Strong absorption from vibrationally excited H$_2$ is also observed toward Herschel 36 (O7.5V, $E_{B-V}$ = 0.87 mag, $f_{H2}$ = 0.03, highlighted as ☆; Rachford et al. 2014), in a cloud apparently very close to the star and its strong UV field (Dahlstrom et al. 2013). The anomalously broad profiles of some of the DIBs (ETR; noted above) are due to a strong local IR radiation field (Dahlstrom et al. 2013; Oka et al. 2013). The presence of ETR in the $\lambda$5780.5 DIB leads to some uncertainties when choosing the continuum region and thus the measured EW of this DIB (although our measurement agrees with the value in Dahlstrom et al. 2013). The $W_\lambda(5780)/W_\lambda(5797)$ ratio is slightly lower than average, for the low $f_{H2}$.

4) The star HD 200775 (B2Ve, $E_{B-V}$ = 0.63 mag, $f_{H2}$ = 0.83, highlighted as ✳) is in front of the reflection nebula NGC 7023 and illuminates several associated photodissociation regions (e.g. Fuente et al. 1990, 1999). The shape of the local radiation field should thus be similar to the field toward HD 37903, given their similar spectral types. It is the only sight line with $f_{H2}$ > 0.8 in our data sample. The normalized EWs for both the $\lambda$5780.5 DIB and (especially) the $\lambda$5797.1 DIB are below the extensions of the general trends in Figures 4E and 4F, yielding the relatively large $W_\lambda(5780)/W_\lambda(5797)$ ratio (for the large $f_{H2}$). The $N$(H) for this sight line is estimated from the weak $\lambda$5780.5 DIB, however, so $N$(H) may be underestimated and $f_{H2}$ overestimated. If the $f_{H2}$ value of this sight line is too high, then all the DIBs measured in this sight line are depressed, which would be very similar to the case of HD 62542 (e.g. Snow et al. 2002; Welty et al., in preparation).

5) The sight line toward HD 73882 (O8V, $E_{B-V}$ = 0.70 mag, $f_{H2}$ = 0.67, highlighted as □) is similar in some respects to the sight line toward HD 37903, where $W_\lambda(5797)$ is ~50% weakened while the $\lambda$5780.5 DIB is of normal strength. On the other hand, unlike the sight line toward HD 37903, relatively strong absorption from various molecular species (e.g. CN, C$_2$, C$_3$, HD, and CO) and relatively weak (but definite) C$_2$ DIBs are detected toward HD 73882 (Oka et al. 2003; Ferlet et al. 2000; Sonnentrucker et al. 2007; this work). Both $N$(K I) and $N$(Na I) are low relative to both $N$(H$_{tot}$) and $N$(H$_2$), suggestive of an enhanced UV field, though probably not one as strong as the field toward HD 37903 (Welty & Hobbs 2001). Snow et al. (2000) also reported a large column density of H$_2$ for $J$ = 4 and 5 levels where the corresponding lines are saturated. Those high column densities are sometimes attributed to the presence of a much stronger than average UV radiation field in the sight line. On the other hand, Lacour et al. (2005) attributed the excitation and broadening of the high-$J$ lines in this sight line to the presence of turbulence in a warm layer in the molecular cloud. Higher resolution spectra will be needed to determine whether the anomalously large $W_\lambda(5780)/W_\lambda(5797)$ ratio seen toward HD 73882 is associated with a stronger than average UV radiation field.

Thus, we have six sight lines (HD 37022, HD 37061, HD 37903, HD 73882, HD 200775, and Herschel 36) where there is some evidence for an enhanced UV radiation field affecting the cloud with the DIBs. Large $W_\lambda(5780)/W_\lambda(5797)$ ratios are found toward five of them, which appears to be due to weakened $W_\lambda(5797)$ rather than enhanced $W_\lambda(5780)$. For the only exception, Herschel 36, we do not see a clear explanation in this context.

4.3 Comparing Behaviors of DIBs to Other Interstellar Species

In this section, we examine the behavior of some known interstellar species at different $f_{H2}$ values, for comparison with the corresponding behavior of the DIBs shown in Figure 4. The normalized column densities of several representative neutral and ionized atomic species (K, Ca, $Ca^+$, O, and $Ti^+$) and of several diatomic molecules (CH, $CH^+$, $C_2$, CO, and CN) versus $f_{H2}$ are given in Figure 7[14]. These trends with increasing $f_{H2}$ reflect the behavior of the column densities of those species versus $N(H)$ and/or $N(H_2)$, as summarized above in Section 4.1. We note that because even the strongest DIBs in our data sample show no evidence of saturation (Friedman et al. 2011), the DIB EWs are thus proportional to the column densities of their respective carriers, and so are directly comparable to the column densities of these known atomic and molecular species.

For the trace neutral forms of little-depleted elements (e.g. K in Figure 7A, and similarly for Na and S, not shown here), the normalized column densities rise monotonically with $f_{H2}$ (with some scatter), reflecting both the ionization behavior of those neutrals and the minimal depletions. The three outliers in the lower right are HD37903, HD200775, and HD37903, already noted as being anomalous in the strength of the $\lambda$5797.1 DIB. For Ca (Figure 7C) and Fe, the normalized column densities exhibit an initial increase, but then decline for $f_{H2} > 0.2$ due to the increasingly severe depletions of calcium and iron in denser gas. $Ca^+$ (Figure 7E), which is a trace species in very diffuse gas but can become dominant (though also more severely depleted) in higher density gas (Welty et al. 1996, 2003), also exhibits an initial rise for small $f_{H2}$ followed by a sharp decline for $f_{H2} > 0.2$. The normalized column densities of the dominant ions of little-depleted elements (e.g. O in Figure 7G, and similarly for $P^+$ and $Zn^+$) are relatively constant for $f_{H2} < 0.1$, but can decline somewhat for higher $f_{H2}$, due to their mild depletions in denser gas (and/or, for $P^+$ and $Zn^+$, to possible underestimation of the column densities from strong absorption lines). The normalized column densities of the dominant ions of typically more severely depleted elements (e.g., $Ti^+$ in Figure 7I and similarly for $Fe^+$) also are roughly constant for small $f_{H2}$, but then decline fairly steeply for $f_{H2} > 0.1$ due to the increasingly severe depletions.

The normalized column densities of the neutral diatomic molecules CH, $C_2$, CO, and CN all rise monotonically with $f_{H2}$ (Figures 7B, F, H, and J). The increase is steeper for the so-called "second-generation" and "third-generation" molecules ($C_2$, CN, and CO), which require other precursor molecules for their formation and which thus trace somewhat denser gas than the "first-generation" CH (e.g., Federman et al. 1994; Pan et al. 2005; see Figure 7B in this work). The normalized column density of CO exhibits an additional steepening of the trend with $f_{H2}$, due to the onset of self-shielding for $f_{H2} > 0.3$ (see Sheffer et al. 2008). The normalized column density of the radical $CH^+$, on the other hand, appears to rise with increasing $f_{H2}$ (for small $f_{H2}$), but then declines for $f_{H2} > 0.3$ (Figure 7D), presumably reflecting its rapid destruction in gas of even relatively modest densities, due to collisions with electrons, H atoms, and/or $H_2$. If the $CH^+$ is produced in turbulent dissipation regions (e.g., Godard et al. 2014), then its decline at higher $f_{H2}$ could also be the result of a reduced formation rate, as such regions are suppressed at higher densities.

---

[14]The column densities used here are largely from Welty & Hobbs 2001; Welty et al. 2003; Sonnentrucker et al. 2007; Jenkins 2009; and Welty & Crowther 2010; an updated compilation is maintained at http://astro.uchicago.edu/~dwelty/coldens.html.

The behavior of most of the known atomic and molecular species thus differs from that of the DIBs considered in this study. While the normalized EWs of the $C_2$ DIBs show little dependence on $f_{H2}$, those of the normal DIBs decline sharply at high $f_{H2}$. For the known atomic and molecular species, however, those whose destruction is thought to be dominated by the radiation field -- via photoionization (the little-depleted trace neutrals) or by photodissociation (the neutral diatomic molecules) -- all exhibit monotonic increases of their normalized column densities with increasing $f_{H2}$. The trace neutral atomic species are never shielded enough, in our sight line sample, to become dominant, however – i.e., the reduced photoionization does not significantly reduce the normalized column densities of the dominant ions with respect to the trace neutrals, even at the highest $f_{H2}$ probed here. The normalized column densities of the dominant ions of little-depleted atomic species (e.g., O, $P^+$) exhibit only modest declines at high $f_{H2}$. The species whose normalized column densities do decline significantly at higher $f_{H2}$ (i.e., Ca, $Ca^+$, $Ti^+$, and $CH^+$; see Figure 7C, E, I, and D) -- more like the behavior shown by the normal DIBs -- all are affected by additional, non-radiative processes that either remove those species from the gas (depletion, collisional destruction) or reduce their formation rates. The turn-over point, where the normalized column densities begin to decline, must reflect the interplay between ionization equilibrium (which dominates in more diffuse gas) and depletion or other destructive processes, which become more significant at higher densities and molecular fractions. These comparisons suggest that ionization and shielding do not account entirely for the observed behavior of the DIBs in diffuse atomic and molecular clouds, and that additional destructive processes are likely to significantly affect the abundances of the DIB carriers in the denser gas characterized by higher molecular fractions ($f_{H2} > 0.3$).

4.4 A Possible Sequence of DIBs

The $C_2$ DIBs defined in Thorburn et al. (2003) may trace the dense parts of interstellar clouds as their normalized EWs do not decline at larger $f_{H2}$, in contrast to the behavior of the normal DIBs. This is clearly seen in Figure 4, as well as in Figure B1 and B2, which give the absolute, unnormalized EWs of the DIBs. The $C_2$ DIBs exist in the sight lines with highest $f_{H2}$ values, which we presume to be the sight lines with highest densities in the clouds.

On the other hand, $W_\lambda(5780)$ is well-correlated with $N(H)$ (Herbig 1993; Friedman et al. 2011), suggesting that its carrier is in the less dense cloud envelopes or in the diffuse inter-cloud gas. The decreasing trend of the $W_\lambda(5780)/W_\lambda(5797)$ ratio with increasing $f_{H2}$ (Figure 5A) implies the carrier of the $\lambda 5797.1$ DIB exists in a denser region (on average) than that of the $\lambda 5780.5$ DIB. Thus, there may be spatial differences in the location of different DIB carriers.

In this section, we propose a more detailed sequence for the eight DIBs studied here according to the behavior of their pair-wise strength ratios as a function of $f_{H2}$ (assuming $f_{H2}$ is a density indicator). In short, if a decreasing trend for W(DIB_A)/W(DIB_B) is observed with increasing $f_{H2}$ (e.g., $W_\lambda(5797)/W_\lambda(4963)$ in Fig. 8A), we conclude that DIB_B favors a higher density region in the diffuse interstellar cloud than DIB_A, and vice versa. Representative plots are given in Figure 8, and we propose the sequence (in increasing order of density of the environment where the carrier of the DIB is found): the $\lambda 6283.8$ and $\lambda 5780.5$ DIBs, the $\lambda 6196.0$ DIB, the $\lambda 6613.6$ DIB, the $\lambda 5797.1$ DIB, and the $C_2$ DIBs. In an onion-like model of interstellar clouds, this proposed sequence would also describe the relative spatial locations of the DIB carriers.

DIBs from the less dense regions should have a better correlation with atomic hydrogen (unless they exist in the ionized edges of diffuse clouds), while DIBs found in denser environments should be better correlated with interstellar species that are also found in the interior of clouds ($H_2$ and CO, for instance). In Tables 2 and 3, following our sequence, the correlation coefficients between $W$(DIB) and $N$(H) are generally increasing, while the correlation coefficients between $W$(DIB) and $N(H_2)$ generally decrease. Moreover, based on the sequence, DIBs in the adjacent portions tend to correlate better. Examples include the moderate correlation between $C_2$ DIBs and the $\lambda$5797.1 DIB, whose carrier is proposed to be found in the denser part of the cloud than any other normal DIB in our sequence. The poor correlations between the $C_2$ DIBs and the $\lambda$5780.5 DIB noted previously (Figure 6) provide another example: those DIBs that are more separated in our sequence are more poorly correlated.

The variation of strength ratio is found in multiple DIB pairs, which have similar behavior to the $W_\lambda(5780)/W_\lambda(5797)$ ratio with increasing $f_{H2}$. The sigma-zeta effect, or the variation in DIB-DIB EW ratio, is thus not restricted to the $\lambda$5780.5 and the $\lambda$5797.1 DIBs. This variation becomes more dramatic when the two DIBs being compared are farther apart in the sequence, and the most dramatic variation in our DIB sample is found between the $C_2$ DIBs and the $\lambda$6283.8 DIB (as given in Figure 8F).

Figure 8 also provides some insights into the well-studied DIBs. For example, Figure 8C indicates that the nearly-perfect correlation between the $\lambda$6196.0 and the $\lambda$6613.6 DIBs breaks down at $f_{H2} <$ 0.15, where more than 90% of the gas is atomic. Most of these sight lines have low reddening; a similar result can be seen in the inset of Figure 5 in McCall et al. (2010) as a turn-over. Galazutdinov et al. (2002) and McCall et al. (2010) note that these two DIBs have different line shapes and widths, and Krełowski et al. (2016) also found that they have different responses to varying interstellar environments. All these evidences suggest that the $\lambda$6196.0 and the $\lambda$6613.6 DIBs are likely due to distinct carriers, despite their very good correlation.

Figure 8E provides a comparison between the $\lambda$5780.5 and the $\lambda$6283.8 DIBs, both of which are known for being comparatively robust when a strong radiation field is present in the sight line. The $W_\lambda(6283)/W_\lambda(5780)$ ratio varies little with $f_{H2}$, with two significant outliers that are Orion Trapezium sight lines (labeled). In those sight lines, the $\lambda$5780.5 DIB is weakened by a factor of five compared to $N$(H) and by two compared to $E_{B-V}$, while the $\lambda$6283.8 DIB is unaffected compared to $E_{B-V}$ (see the full version of Table 1). The densities of UV photons in the Trapezium sight lines are greatly enhanced by the proximate O-type stars (Krełowski and Sneden 1995), and the carrier of the $\lambda$6283.8 DIB seems to be more robust in such environments.

There have been several sequences of DIBs proposed in the literature. For example, Welty (2014) proposed a sequence based on the slopes between $W$(DIB) and the column densities of known interstellar species [e.g., $N$(H), $N(H_2)$, $N$(CH), $N(CH^+)$, $N(H_{tot})$, $E_{B-V}$, $N$(Na), $N$(K), and $N$(Ca)]. For the six DIBs we have in common (DIBs $\lambda\lambda$4963.9, 5780.5, 5797.1, 6196.0, 6283.8, and 6613.6), the sequences agree with each other, except for the $\lambda$6613.6 DIB. Sonnentrucker et al. (1997) discussed the ionization potentials of the carriers of four DIBs, yielding a sequence consistent with the one proposed above. Additional support for this sequence comes from Lan et al. (2015) who studied the distribution of DIBs using low-resolution SDSS data (York et al. 2000). They proposed a dependence of $W$(DIB) on the column densities of H and $H_2$:

$$W(\text{DIB}) = W_{21} * (\frac{N(\text{H})}{10^{21}})^\alpha * (\frac{N(\text{H}_2)}{10^{21}})^\mu$$

Following the sequence of DIBs we propose, $\mu$ shows an increasing trend (see Table 4 and Figure 13 of Lan et al. 2015). Such a trend indicates a growing dependence on $H_2$ from DIBs $\lambda\lambda$5780.5 and 6283.8 to the $C_2$ DIBs in our sequence. Although these studies are based on different methods, the overall trends are in good agreement with each other, apparently reflecting properties of the carriers of these DIBs.

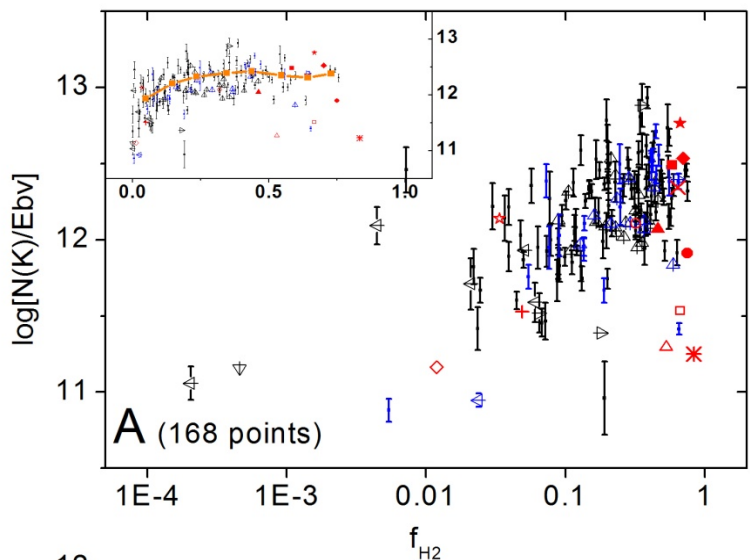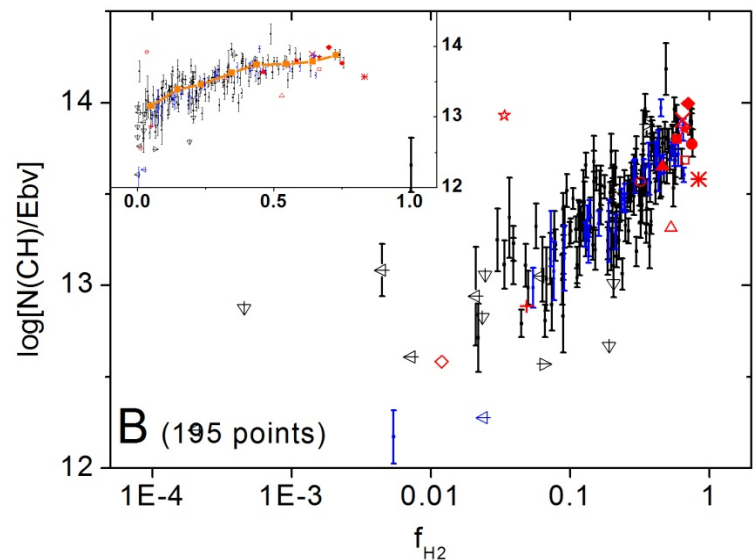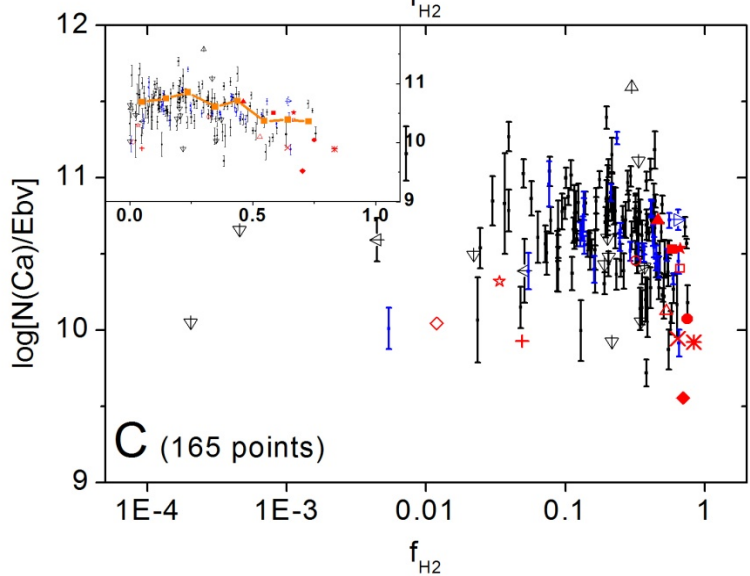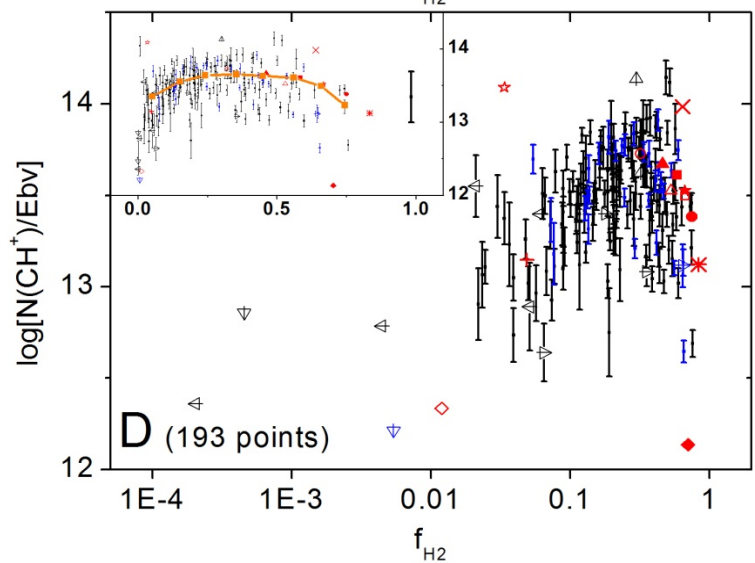

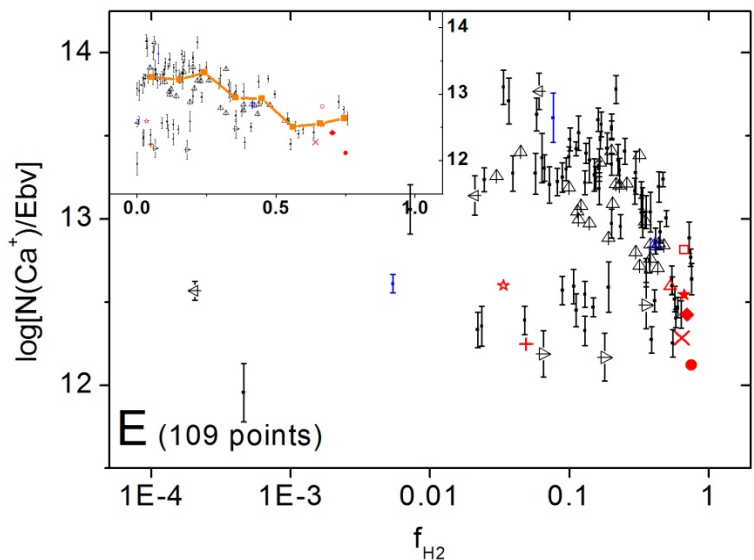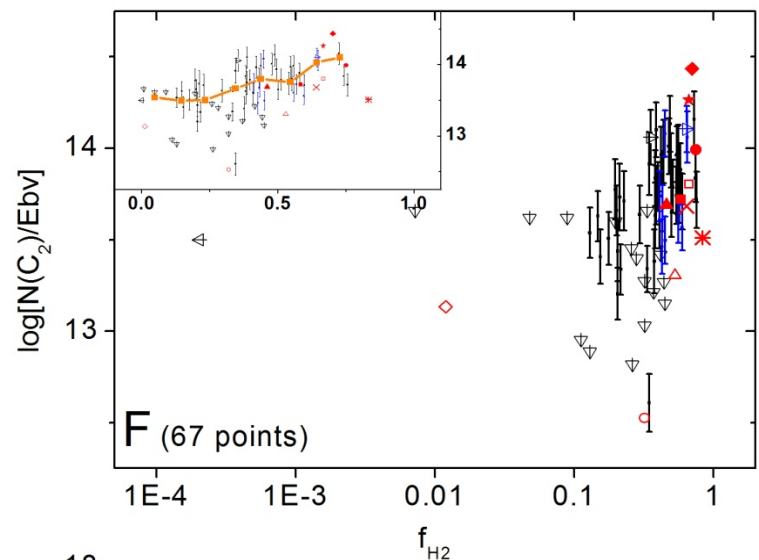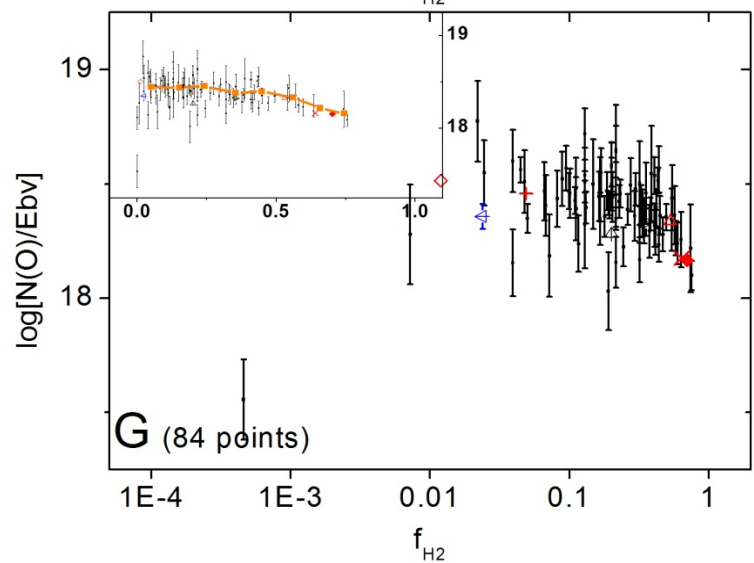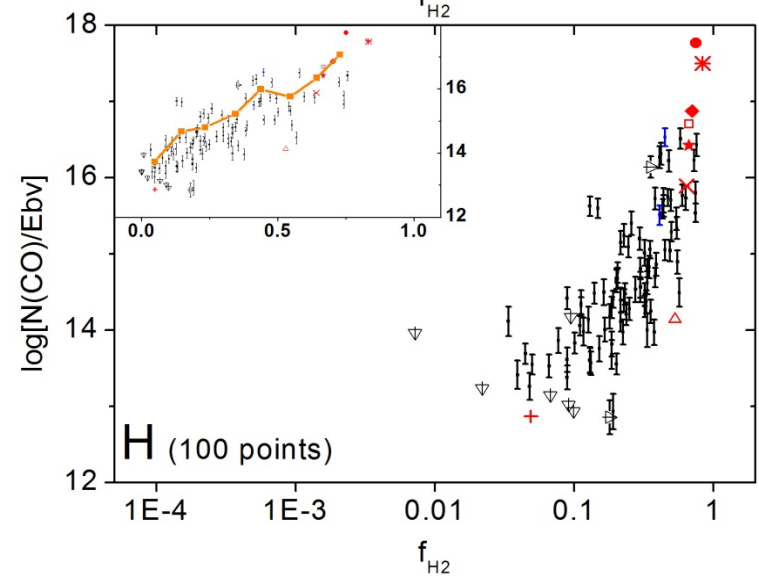

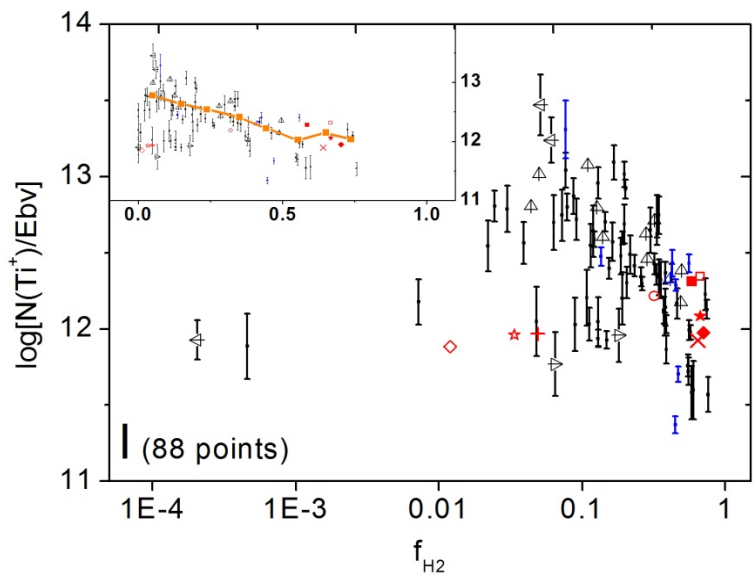
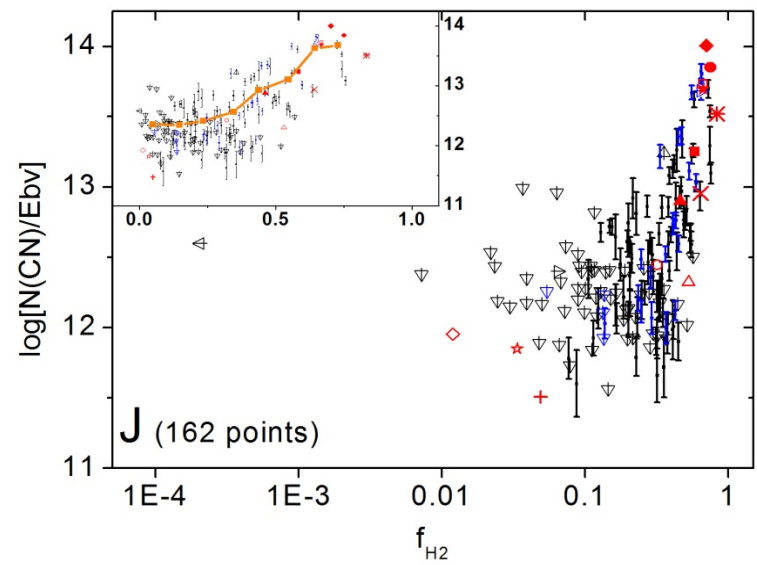

Figure 7. Behavior of some known interstellar species as a function of $f_{H2}$. The ordinates in the main panels are logarithmic ratios of the column densities of the species divided by $E_{B-V}$. The ordinates of the insets in each panel are the numerical values of the same ratios (please compare to the format of Figure 4). The symbolic system used here is the same as in Figure 4, where black points are sight lines with directly measured $f_{H2}$ values while points based on the use of surrogates are plotted as blue points. Arrows toward different directions indicate upper or lower limits for $f_{H2}$ and/or normalized column densities of different species, and orange dots and lines in the inset of each panel give the average of each bin. Unlike normal DIBs, the decrease at $f_{H2} > 0.3$ is not found for all species. For those species with decreases at relatively large $f_{H2}$ value, the decreases are likely related to depletion onto grains in the denser parts of the interstellar cloud (i.e. Ca, $Ca^+$, and $Ti^+$) or collisional destruction (i.e. $CH^+$). These plots are based on a more extended data sample than the sample used for DIB study.

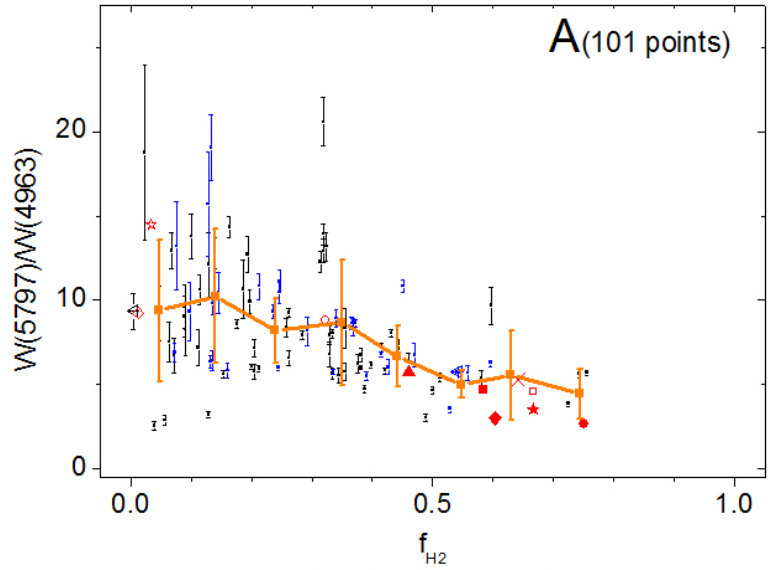
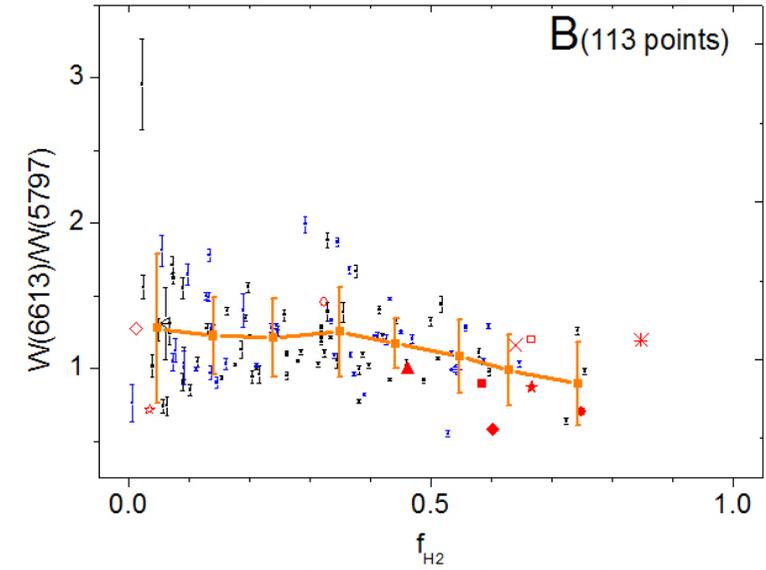
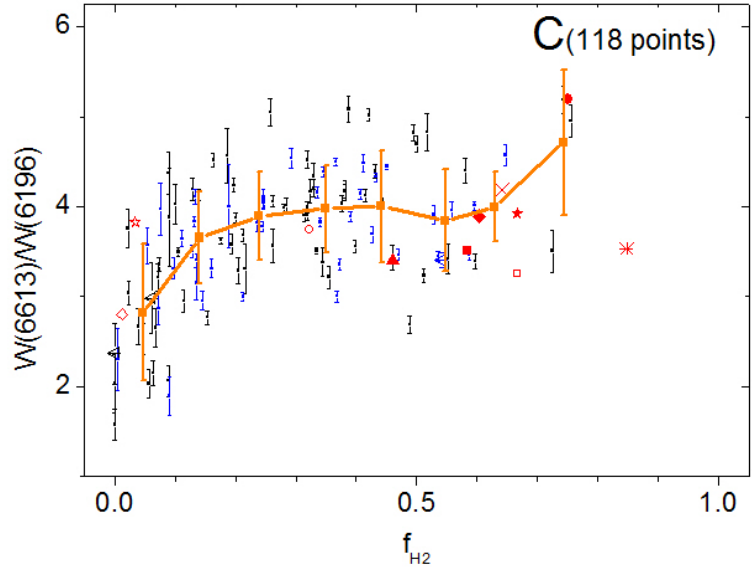
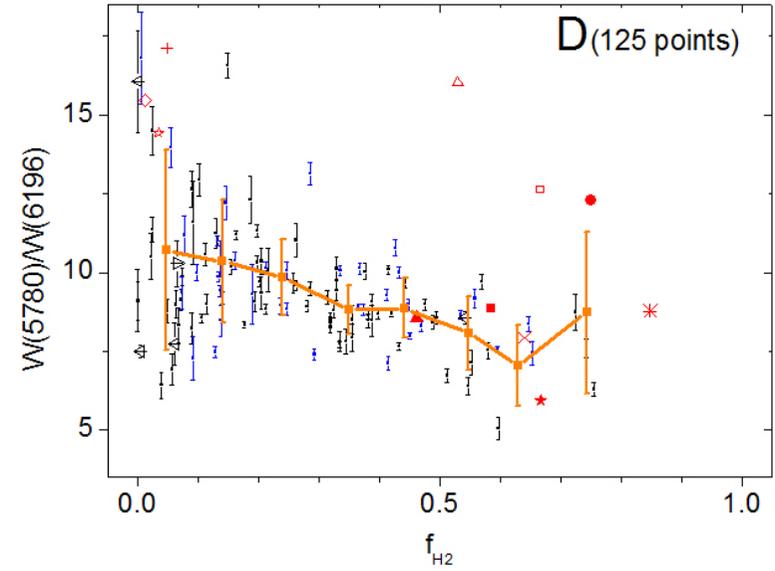

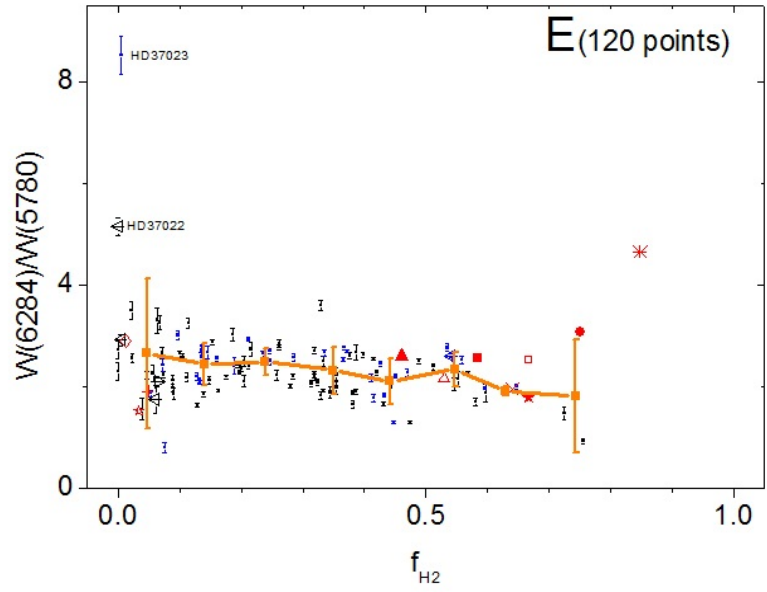
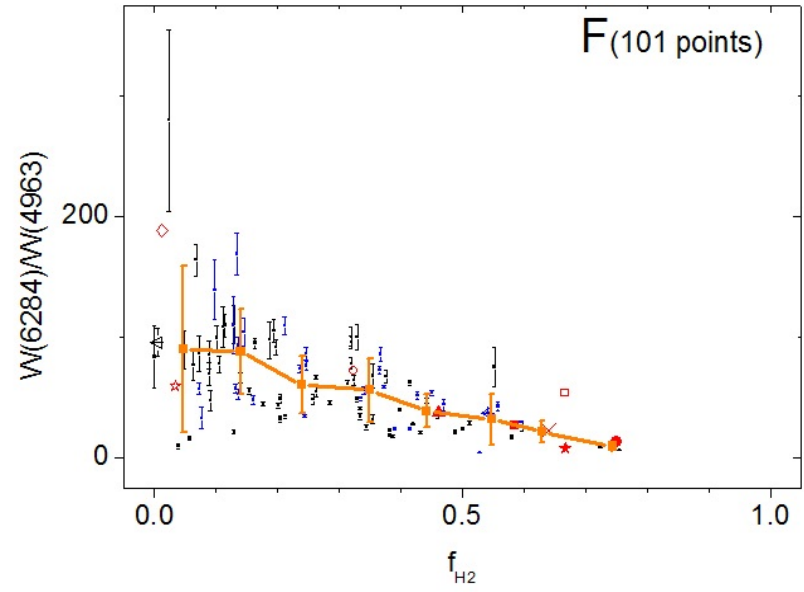

Figure 8. Behavior of DIB-DIB ratios as a function of $f_{H2}$. As explained in the text, we assume that a negative slope indicates the DIB in the denominator of the ordinate of each plot is in a higher density region of interstellar clouds than the DIB in the numerator. The sequence of the DIBs in increasing order of density of environment therefore is: the $\lambda 6283.8$ and $\lambda 5780.5$ DIBs, the $\lambda 6196.0$ DIB, the $\lambda 6613.6$ DIB, the $\lambda 5797.1$ DIB, and the $C_2$ DIB $\lambda 4963.9$.

## 5. CONCLUSIONS

We used high quality spectra of 186 Galactic sight lines to explore the behavior of five normal DIBs ($\lambda\lambda$5780.5, 5797.1, 6196.0, 6283.8, and 6613.6) and three $C_2$ DIBs ($\lambda\lambda$4726.8, 4963.9, and 4984.8) in diffuse atomic and molecular cloud environments, as characterized by the molecular fraction $f_{H2}$. Because the DIBs are so broad compared to known atomic and molecular lines, and not readily assignable to interstellar components as defined by high resolution interstellar K I observations, we use integrated sight line values for the normalized equivalent widths of DIBs [$W$(DIB)/$E_{B-V}$, referred to as normalized EWs]; $f_{H2}$ (assumed to be a density indicator); the color excess, $E_{B-V}$; and the integrated column densities of selected interstellar atomic and molecular species, $N$(X). The $E_{B-V}$ in our data sample ranges from 0.00 to 3.31 mag, and we have 144 direct or surrogate measurements of $f_{H2}$ ranging from less than $10^{-4}$ to 0.83. Each plot comparing normalized EWs of DIBs (or their ratios) with $f_{H2}$ typically contains measurements from about 100 sight lines. Compared to similar studies (e.g. Sonnentrucker et al. 1997, Cami et al. 1997, Vos et al. 2011), we have increased the target sample by at least a factor of 2, particularly for more reddened sight lines.

We have reached the following conclusions.

1. The normalized equivalent widths of the normal DIBs increase with increasing $f_{H2}$ for $f_{H2}$ < 0.15, and decrease for $f_{H2}$ > 0.3. We refer to this behavior as a Lambda-shaped behavior. The different DIBs exhibit different slopes in the increasing and decreasing regimes and slightly different locations of the peak $f_{H2}$ value. The peak between the two slopes and the decline at higher $f_{H2}$ is not evident for the $C_2$ DIBs. All DIBs measured here exhibit cosmic scatter, in the sense that the dispersion of data points at any given $f_{H2}$ is much larger than the uncertainties in the individual measurements. The term "Lambda-shaped behavior" is meant to describe the general trend of the envelope.

2. The ratio of $W_\lambda$(5780)/$W_\lambda$(5797) varies over a factor of 5 in our data sample, which is known as the sigma-zeta effect. By comparing this ratio to $f_{H2}$, we confirm that the highest ratios only occur at the lowest values of $f_{H2}$, except for three outliers (HD 37903, HD 73882, and HD 200775). Some of these outlying measurements involve interstellar clouds that could lie near the background stars, suggesting that the radiation field may play a role in the DIB behaviors. While there is ambiguity in how to draw the continuum of the $\lambda$5797.1 DIB, we show in Appendix A that our results are independent of this issue.

3. The normalized column densities of four neutral diatomic interstellar molecules (CH, CO, $C_2$, and CN), and of the trace neutral forms of little-depleted elements (e.g. K) increase monotonically with $f_{H2}$ -- unlike the behaviors of the five normal DIBs, with their Lambda-shaped behavior, and the three $C_2$ DIBs, which show the flatter behavior noted earlier, most notably at large $f_{H2}$. A decline at large $f_{H2}$ is observed for the normalized column densities of Ca, $Ca^+$, $Ti^+$ (and other trace and dominant forms of typically depleted elements), and $CH^+$. The declines for those species at higher densities are attributed to additional destructive processes, such as condensation onto grains, various collisional processes, or depressed formation rate. Such additional, non-radiative processes -- rather than the usually assumed shielding effects -- may be responsible for the observed declines in the normalized EWs of the "normal" DIBs for $f_{H2}$ > 0.3.

4. We propose a sequence of DIBs based on their dependences on $f_{H2}$ (and thus on density) in an increasing order of density within the interstellar clouds: the $\lambda$6283.8 and $\lambda$5780.5 DIBs, the $\lambda$6196.0 DIB, the $\lambda$6613.6 DIB, the $\lambda$5797.1 DIB, and the $C_2$ DIBs $\lambda\lambda$4726.8, 4963.9 and 4984.8. This ordering presumably reflects local properties that affect the DIBs, e.g. robustness to destruction in ambient radiation fields and/or particular densities favored by the carrier of each DIB. This sequence largely agrees with those suggested by Welty (2014) (except for the $\lambda$6613.6 DIB), Lan et al. (2015) and Sonnentrucker et al. (1997), all of which are based on different methods.

5. Using a larger data set than in Thorburn et al (2003) ($C_2$ DIBs detected toward 154 stars compared to their 18 detections toward 35 stars), we show that good correlations exist among three of the $C_2$ DIBs: $\lambda\lambda$4726.8, 4963.9, and 4984.8, and that they are better correlated with $N(H_2)$ than with $N(H)$. The $C_2$ DIBs do not correlate well with normal DIBs, consistent with their belonging to a distinct DIB class.

HF and GZ are supported by the National Natural Science Foundation of China under grant Nos. 11390371, 11233004, and 11603033. DGY was supported by NSF grant AST-1009603, DEW by NSF grant AST-1238926, JAD by NSF grant AST-1008424, TPS by NSF grant AST-1009929, BLR by NSF grant AST-1008801.
Based in part on data obtained from the ESO Science Archive Facility by user "zihaoj".

*Facilities*: APO (ARCES), ESO/Max Planck: 2.2m (FEROS), Magellan: Clay (MIKE)

**REFERENCES**

Adamson, A. J., Whittet, D. C. B., & Duley, W. W. 1991, MNRAS, 252, 234
Bernstein, R., Shectman, S. A., Gunnels, S. M., et al. 2003, Proc. SPIE, 4841, 1694
Bohlin, R. C., Savage, B. D., & Drake, J. F. 1978, ApJ, 224, 132
Cami, J., Sonnentrucker, P., Ehrenfreund, P., et al. 1997, A&A, 305, 616
Campbell, E.K., Holz, M., Gerlich, D., & Maier, J. P. 2015, Nature, 523, 322
Cardelli, J. A. 1994, Science, 265, 209
Cardelli, J. A., & Savage, B. D. 1988, ApJ, 325, 864
Code, A. D. 1958, PASP, 70, 261
Cordiner, M. A., Fossey, S. J., Smith, A. M., & Sarre, P. J. 2013, ApJ, 764, 10
Cox, N. L. J., Cordiner, M. A., Cami, J. et al. 2006, A&A, 447, 991
Cox, N. L. J., Cordiner, M. A., Ehrenfreund, P. et al. 2007, A&A, 470, 941
Crane, P., Lambert, D. L., & Sheffer, Y. 1995, ApJS, 99, 107
Dahlstrom, J., York, D. G., Welty, D. E., et al. 2013, ApJ, 773, 41
Danks, A. C., Dennefeld, M., Wamsteker, W., & Shaver, P. A. 1983, A&A, 118, 301
Douglas, A. E., & Herzberg, G. 1941, ApJ, 94, 381
Draine, B. T., & Bertoldi, F. 1996, ApJ, 468, 269
Duley, W. W. 2006, Faraday Discuss., 133, 415
Ensor, T., Cami, J., Bhatt, N., & Soddu, A. 2017, ApJ, 836, 162
Federman, S. R., Sneden, C., Schempp. W. V., & Smith, W. H. 1985, ApJ, 290, 55
Federman, S. R., Strom, C. J., Lambert, D. L. et al. 1994, ApJ, 424, 772
Ferlet, R., Andre, M., Hebrand, G., et al. 2000, ApJ, 538, 69
Fitzpatrick, E., & Massa, D. 1990, ApJ, 72, 163
Foing, B. H., & Ehrenfreund, P. 1994, ApJ, 369, 296
Friedman, S. D., York, D. G., McCall, B. J., et al. 2011, ApJ, 727, 33
Fuente, A., Martin-Pintado, J., Bachiller, R., & Cernicharo, J. 1990, A&A, 237, 471
Fuente, A., Martin-Pintado, J., Rodriguez-Fernandez, N. J., et al. 1999, ApJ, 518, 45
Galazutdinov, G. A., Lo Curto, G., & Krelowski, J. 2008, MNRAS, 386, 2003
Galazutdinov, G. A., Manico, G., & Pirronello, V., et al. 2004, MNRAS, 355, 169
Galazutdinov, G. A., Moutou, C., Musaev, F., & Krełowski, J. 2002, A&A, 384, 215
Galazutdinov, G. A., Shimansky, V. V., Bondar, A., et al. 2017, MNRAS, 465, 3956
Glassgold, A. E., Mamon, G. A., Omont, A., & Lucas, R. 1987, A&A, 180, 183
Gnaciński, P. 2011, A&A, 532, 122

Godard, B., Falgarone, E., & Pineau des Forets, G. 2014, A&A, 570, 27
Gredel, R., Carpentier, Y., Rouillé, G., et al. 2011, A&A, 530, 26
Guthe, F., Tulej, M., Pachkov, M. V., & Maier, J. P. 2001, ApJ, 649, 299
Heger, M. L. 1922, Lick Obser. Bull., 10, 141
Herbig, G. H. 1993, ApJ, 407, 142
Herbig, G. H. 1995, ARA&A, 33, 19
Hobbs, L. M., York, D. G., Snow, T. P. et al. 2008, ApJ, 680, 1256
Hobbs, L. M., York, D. G., Thorburn, J. A., et al. 2009, ApJ, 705, 32
Huang, J., & Oka, T. 2015, MolPh, 113, 2159
Jenkins, E. B., Woźniak, P. R., Sofia, U. J., et al. 2000, ApJ, 538, 275
Jenkins, E. B. 2009, ApJ, 700, 1299
Jenniskens, P., Ehrenfreud, P., & Foing, B. 1994, A&A, 281, 517
Jensen, A. G., & Snow, T. P. 2007a, ApJ, 669, 378
Jensen, A. G., & Snow, T. P. 2007b, ApJ, 669, 401
Jensen, A. G., Snow, T. P., Sonneborn, G., & Rachford, B. L. 2010, ApJ, 711, 1236
Johnson, H. L. 1963, in Stars and Stellar Systems 3, Basic Astronomical Data, ed. K. A. Strand (Chicago: Univ. Chicago Press), 214
Josafatsson, K., & Snow, T. P. 1987, ApJ, 319, 436
Kaufer, A., Stahl, O., Tubbesing, S., et al. 2000, Proc. SPIE, 4008, 459
Kerr, T. H., Hibbins, R. E., Fossey, S. J., et al. 1998, ApJ, 495, 941
Kos, J., & Zwitter, T. 2013, ApJ, 774, 72
Krelowski, J., Ehrenfreund, P., Foing, B. H. et al. 1999, A&A, 347, 235
Krełowski, J., Galazutdinov, G. A., Bondar, A., & Beletsky, Y. 2016, MNRAS, 460, 2706
Krełowski, J., Schmidt, M., & Snow, T. P. 1997, PASP, 109, 1135
Krełowski, J., & Sneden, C. 1995, in The Diffuse Interstellar Bands, ed. A. G. G. M. Tielens, & T. P. Snow (Dordrecht: Kluwer), 13
Krełowski, J., Snow, T. P., Seab, C. G., & Papaj, J. 1992, MNRAS, 258, 693
Krełowski, J., & Walker, G. A. H. 1987, ApJ, 312, 860
Krełowski, J., & Westerlund, B. E. 1988, A&A, 190, 339
Kwok, S., & Zhang, Y. 2013, ApJ, 771, 5
Lacour, S., Ziskin, V., Hébrand, G., et al. 2005, ApJ, 627, 251
Lakin, N. M., Pachkov, M., Tulej, M., et al. 2000, J. Chem. Phys., 113, 9586
Lan, T. W., Menard, B., & Zhu, G. 2015, MNRAS, 452, 3629
Maier, J. P., Walker, G. A. H., & Bohlender, D. A. 2002, ApJ, 566, 332
Maier, J. P., Walker, G. A. H., & Bohlender, D. A. 2004, ApJ, 602, 286
McCall, B. J., Drosback, M. M., Thorburn, J. A., et al. 2010, ApJ, 708, 1628
McCall, B. J., & Griffin, R. E. 2011, RSPSA, 469, 9641
McCall, B. J., Thorburn, J., & Hobbs, L. M. et al. 2001, ApJ, 559, 49
Merrill, P. W. 1934, PASP, 46, 206
Merrill, P. W. 1936, PASP, 48, 179
Meyer, D., Lauroesch, J. T., Sofia, U. J., et al. 2001, ApJ, 553, 59
Meyer, D. M., & Ulrich, R. K. 1984, ApJ, 283, 98
Morton, D. C. 1975, ApJ, 197, 85
Morton, D. C. 2003, ApJS, 149, 205
Motylewski, T., Linnartz, H., Vaizert, O., et al. 2000, ApJ, 531, 312
Moutou, C., Krełowski, J., D'Hendecourt, L., & Jamroszczak, J. 1999, A&A, 351, 680
Oka, T., & McCall, B. J. 2011, Sci, 311, 293
Oka, T., Thorburn, J. A., McCall B. J., et al. 2003, ApJ, 582, 823
Oka, T., Welty, D. E., Johnson, S., et al. 2013, ApJ, 773, 42
Omont, A. 2016, A&A, 590, 52

Pan, K., Federman, S. R., Sheffer, Y., & Andersson, B. G. 2005, ApJ, 633, 986
Rachford, B. L., Snow, T. P., Destree, J. D., et al. 2009, ApJS, 180, 125
Rachford, B. L., Snow, T. P., Tumlinson, J., et al. 2002, ApJ, 577, 221
Rouan, D., Leger, A., & Le Coupanec, P. 1997, A&A, 324, 661
Ruiterkamp, R., Cox, N. L. J., Spaans, M., et al. 2005, A&A, 432, 515
Rachford, B. L., Snow, T. P., & Ross, T. L. 2014, ApJ, 786, 159
Salama, F., Galazutdinov, G. A., Lrelowski, J., et al. 2011, ApJ, 728, 154
Sarre, P. J., Miles, J. R., Kerr, T. H., et al. 1995, MNRAS, 277, 41
Savage, B. D., Bohlin, R. C., Drake, J. F., & Budich, W. 1977, ApJ, 216, 291
Sellgren, K., Werner, M. W., & Dinerstein, H. L. 1983, ApJ, 271, 13
Sembach, K. R., & Savage, B. D. 1992, ApJS, 83, 147
Sheffer, Y., Rogers, M., Federman, S. R., et al. 2008, ApJ, 687, 1075
Sneden, C., Woszczyk, A., & Krełowski, J. 1991, PASP, 103, 1005
Snow, T. P., & Cohen, J. G. 1974, ApJ, 194, 313
Snow, T. P., & McCall, B. J. 2006, ARA&A, 44, 367
Snow, T. P., Rachford, B. L., Tumlinson, J., et al. 2000, ApJ, 538, 65
Snow, T. P., Welty, D. E., Thorburn, J., et al. 2002, ApJ, 573, 670
Solomon, P. M., & Wickramasinghe, N. C. 1969, ApJ, 158, 449
Sonnentrucker, P., Cami, J., Ehrenfreund, P., & Foing, B. H. 1997, A&A, 327, 1215
Sonnentrucker, P., Foing, B. H., Breitfellner, M., & Ehrenfreund, P. 1999, A&A, 346, 936
Sonnentrucker, P., Welty, D. E., Thorburn, J. A., & York, D. G. 2007, ApJS, 168, 58
Strom, K. M., Strom, S. E., Carrasco, L., & Vrba, F. J. 1975, ApJ, 196, 489
Swings, P. 1937, MNRAS, 97, 212
Swings, P., & Rosenfeld, L. 1937, ApJ, 86, 483
Thorburn, J. A., Hobbs, L. M., McCall, B. J., et al. 2003, ApJ, 584, 339
Tulej, M., Kirkwood, D. A., Pachkov, M., & Maier, J. P. 1998, ApJL, 506, L69
Vos, D. A. I., Cox, N. L. J., Kaper, L., et al. 2011, A&A, 533, 129
Voshchinnikov, N. V., Il'in, V. B., Henning, Th., & Dubkova, D. N. 2006, A&A, 445, 167
van Dishoeck, E. F., Black, J. H. 1988, ApJ, 334, 771
Walker, G. A. H., Bohlender, D. A., Maier, J. P., & Campbell, E. K. 2015, ApJ, 812, 8
Wampler, E. J. 1966, ApJ, 144, 921
Wang, S., Hildebrand, R. H., Hobbs, L. M., et al. 2003, Proc. SPIE, 4841, 1145
Warin, S., Benayoun, J. J., & Viala, Y. P. 1996, A&A, 306, 935
Welty, D. E. 2014, IAUS, 297, 153
Welty, D. E., & Crowther, P. A. 2010, MNRAS, 404, 1321
Welty, D. E., Federman, S. R., Gredel, R., et al. 2006, ApJS, 165, 138
Welty, D. E., & Hobbs, L. M. 2001, ApJS, 133, 345
Welty, D. E., Hobbs, L. M., & Kulkarni, V. P. 1994, ApJ, 436, 152
Welty, D. E., Hobbs, L. M., & Morton, D. C. 2003, ApJS, 147, 61
Welty, D. E., Morton, D. C., & Hobbs, L. M. 1996, ApJS, 106, 533
Welty, D. E., Ritchey, A. M., Dahlstrom, J. A., & York, D. G. 2014. ApJ, 792, 106
Weselak, T., Fulara, J., Schmidt, M.R., & Krełowski, J. 2001, A&A, 377, 677
Weselak, T., Galazutdinov, G. A., Musaev, F. A., & Krelowski, J. 2004, A&A, 414, 949
Weselak, T., Galazutdinov, G. A., Musaev, F. A., & Krelowski, J. 2008, A&A, 484, 381
Westerlund, B. E., & Krełowski, J. 1989, A&A, 218, 216
York, D. G., Adelman, J., Anderson, J. E. Jr., et al. 2000, AJ, 120, 1579
Zhen, J., Castellanos, P., Paardekooper, D. M., et al. 2014, ApJL, 797, L30
Zsargó, J., & Federman, S. R. 2003, ApJ, 589, 319

# APPENDIX A. A CLOSER LOOK AT THE λ5797.1 DIB

The presence of the λλ5795.2 and 5793.2 DIBs makes it hard to determine the correct continuum for the λ5797.1 DIB. We use the sketch in Figure A1 to demonstrate typical continua adopted in different works. For instance, the measuring technique adopted by Galazutdinov et al. (2004), 5797-G in Figure A1, uses only the blue region for the EW of the λ5797.1 DIB. This implies the assumption that there is a shallow, broad DIB (red and green) that is about 4Å wide. On the other hand, Thorburn et al. (2003) and Friedman et al. (2011) took the λ5795.2 DIB as part of the λ5797.1 DIB, thus including all the three colored regions (5797-T in Figure A1). The uncertainty introduced from cutting-off the λ5793.2 DIB was small given its typically shallow profile. We choose to include the green region as part of the λ5797.1 DIB (5797-F in Figure A1) to draw the most reproducible continuum, while not combining the λλ5797.1 and 5795.2 DIBs because the profiles in the two atlas stars, HD 183143 and HD 204827 indicate the two DIBs are not correlated (Figure 1). This is a somewhat arbitrary decision, but the finding below suggests the behavior of the λ5797.1 DIB is not sensitive to the continuum chosen.

Until we understand the true profile of the λλ5795.2 and 5797.1 DIBs, any attempt to separate these two DIBs could add errors in the measurements. For the time being, there is limited information on their profiles, particularly the red wing of the λ5795.2 DIB (the green region in Figure A1). A larger sample of higher resolution measurements of this region will probably be needed before the correct continuum for the λ5797.1 DIB can be drawn. Alternatively, a laboratory identification of any of the three features would provide guidance in proper measurement. This type of consideration applies to many DIB profiles: while the formal uncertainties in DIB EWs can be arbitrarily small, such systematic uncertainties can lead to difficulty in comparing correlation coefficients between observers.

However, the $W_\lambda(5797)$ from the different measuring techniques discussed above show similar behavior in dense cloud environments (Figure A2). Their general characteristics are no different from each other, which is likely a result of the fact that the λ5797.1 feature (the blue region in Figure A1) simply dominates this region. We thus emphasize that the behavior of the λ5797.1 DIB is not sensitive to the choice of continuum, as far as can be determined.

The λ5797.1 DIB contains multiple substructures in ultra-high resolution spectra (Sarre et al. 1995). These substructures can be due to different branches of a single molecule (Oka et al. 2013; Huang & Oka 2015), or they may represent blending from DIBs from different carriers that happen to be at similar wavelengths. If these substructures are due to DIBs with different properties, such as a $C_2$ DIB and a normal DIB, the sigma-zeta effect could still be observed given the different behavior of $C_2$ DIBs and normal DIBs in dense cloud environments (Figures 5 and 8).

We therefore searched for a possible $C_2$ DIB component in the residual of the λ5797.1 DIB, $W_{residual}(5797) = W_\lambda(5797) - \alpha*W_\lambda(5780)$. The index $\alpha$ here represents the relative strength of this hypothesized normal DIB and should be the minimum $W_\lambda(5797)/W_\lambda(5780)$ ratio observed (when the $C_2$ DIBs are absent in the sight line), which is 0.121 in this study. We set $\alpha$ from 0.0 to 0.3 and compared the $W_{residual}(5797)$ at different $\alpha$ to $W_\lambda(4963)$, a representative of the $C_2$ DIBs.

The best correlation between $W_{residual}(5797)$ and $W_\lambda(4963)$ is found when $\alpha = 0.17$, where the correlation coefficient is 0.873. This correlation is comparable to the correlations among $C_2$ DIBs (Tables 2 and 3, Figure 6, see also Figure A3). In this case, the hypothesized $C_2$ DIB would be about half the strength of the λ4726.8 DIB. However, the value of chi-square for this correlation is 32.3, significantly larger than that of correlations among the $C_2$ DIBs. Sight lines known for containing strong $C_2$ DIBs (highlighted) contribute a lot to the scatter, as they are usually far above the best-fit line. This implies that the behavior of $W_{\text{residual}}(5797)$ is not totally consistent with the C$_2$ DIBs, especially in extreme cases.

Our attempt to separate a C$_2$ DIB from the $\lambda$5797.1 DIB was thus unsuccessful. This agrees with the finding that substructures in the $\lambda$5797.1 DIB can be reproduced by a single molecule (Oka et al. 2013; Huang & Oka 2015). This conjecture can be tested by comparing the relative strengths between the multiple components of the $\lambda$5797.1 DIB in different sight lines, at higher resolution than is currently generally available. As shown in this work, the C$_2$ DIBs and normal DIBs have different behaviors in different density regions in diffuse atomic and molecular clouds, which can be used to indicate if those structures are DIBs with different properties.

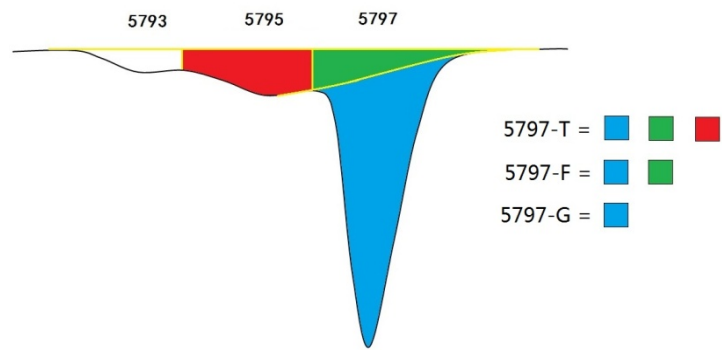

Figure A1. A sketch of different measurement techniques for the $\lambda$5797.1 DIB adopted in different work.

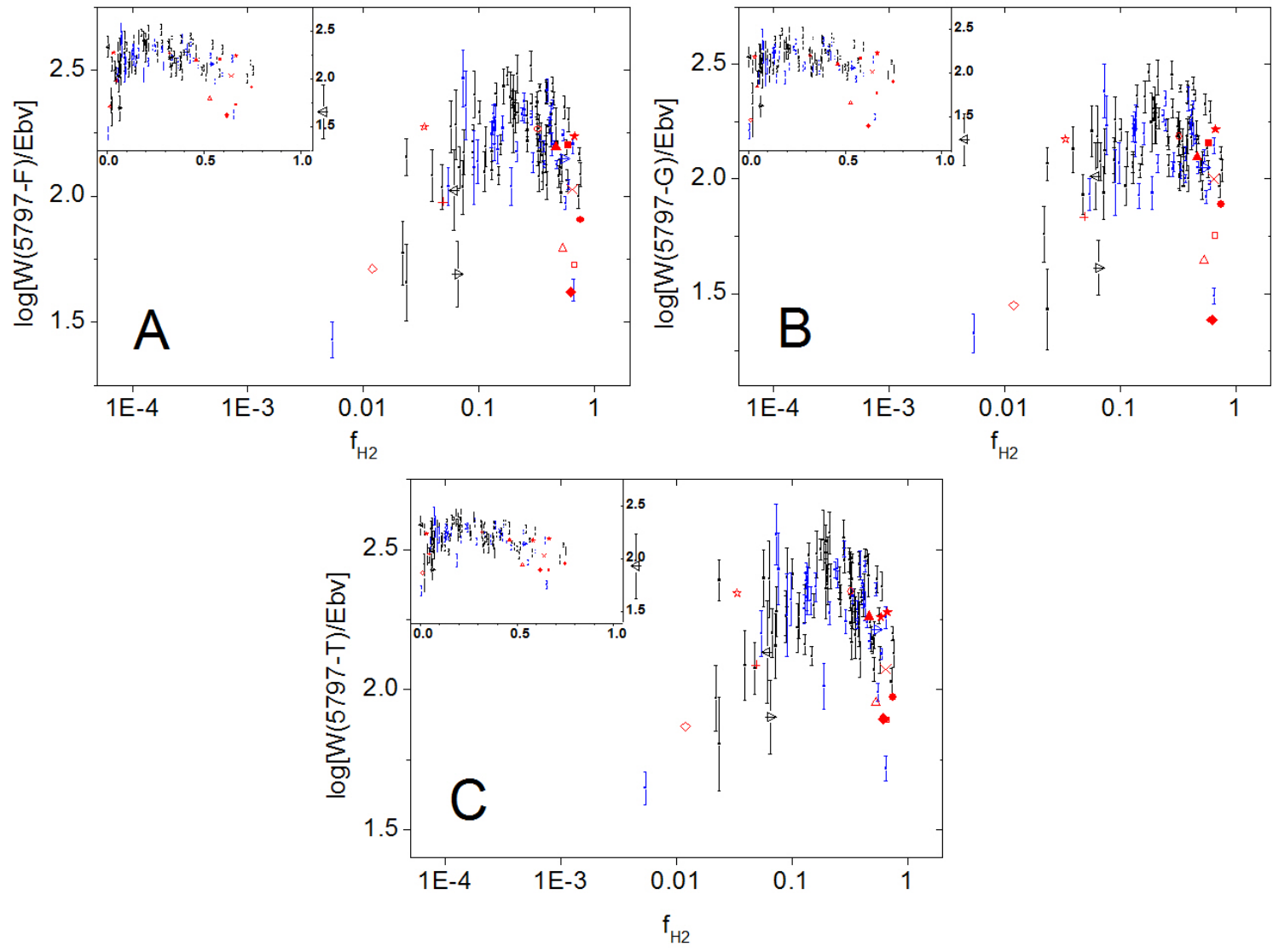

Figure A2. Normalized EWs of the $\lambda$5797.1 DIB from different definitions of the local continuum, as functions of $f_{H2}$, showing that the general Lambda-shaped behavior is the same for the three sets of measurements. Each panel contains measurements from 108 sight lines. The inset of each plot represents the same data with the abscissa plotted on a linear scale.

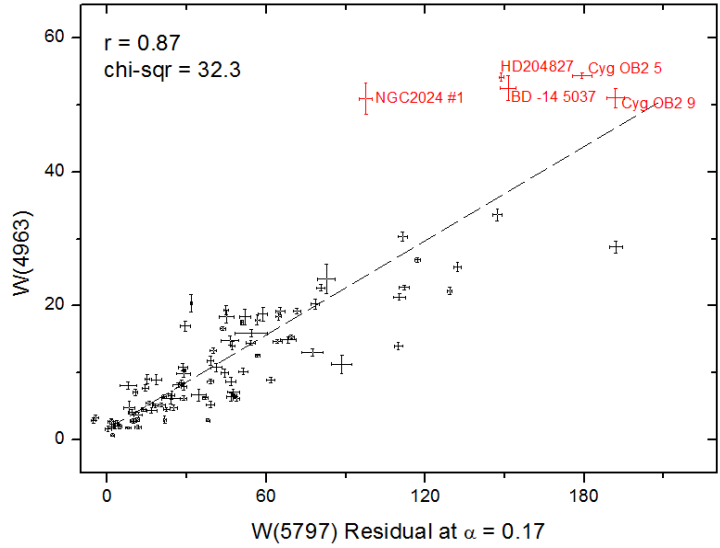

Figure A3. Correlation between residual of $W_\lambda(5797)$ at $\alpha = 0.17$ and the $\lambda 4963.9$ $C_2$ DIB. The correlation coefficients and chi-square given in the plot are obtained as described in Section 3.5. Note the larger scatter compared to the correlations among the $C_2$ DIBs. We highlighted some sight lines where exceedingly strong $C_2$ DIBs are detected. They are usually far above the best-fit line and are inconsistent with the $\lambda 5797.1$ DIB containing a $C_2$ DIB component.

# APPENDIX B. BEHAVIOR OF UNNORMALIZED DIBS AT DIFFERENT $f_{H2}$

In Figure B1, we present the behavior of *unnormalized* DIBs with the same symbol-system used in Figure 4. The Lambda-shaped behavior is still observed for all the five normal DIBs, and the three $C_2$ DIBs undergo more obvious growth throughout the $f_{H2}$ range. Compared to Figure 4, the trends shown in Figure B1 have larger scatter, especially at $f_{H2} > 0.2$. This larger scatter is likely the result of averaging multiple clouds along the sight line and of including a mixture of stars of different distance and reddening (see also Figure B2), and motivates our use of $E_{B-V}$ to normalize the EWs of DIBs in section 3.2.

Figure B2 provides a comparison of the relative strengths of each DIB in each sight line. Each measurement is normalized by the maximum EW of that DIB observed in this work, so that the ordinate of each panel ranges between 0 and 1. Vertical lines are used to line up the points in each panel from the same sight line. The data points are colored based on the distance of the background star, where red is the nearest and black is the most distant, and the special symbols are kept for the sight lines of interest as in Figure 4.

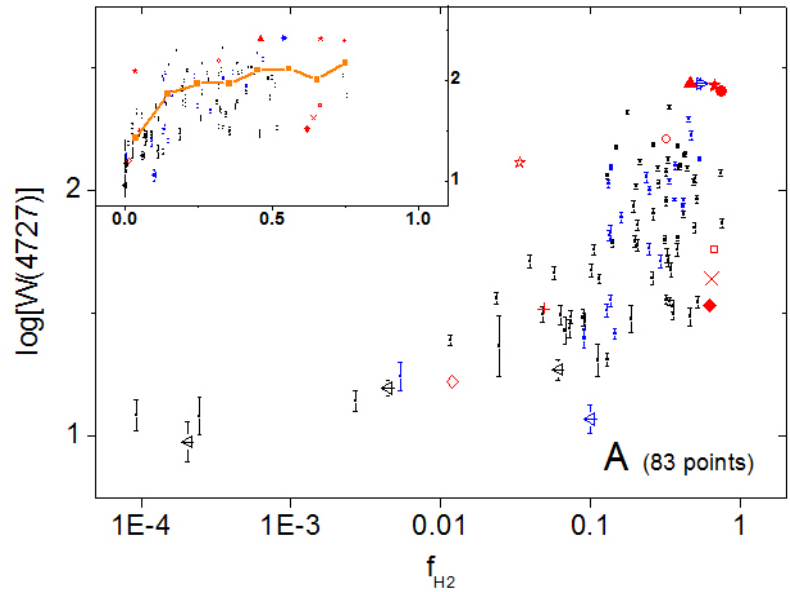
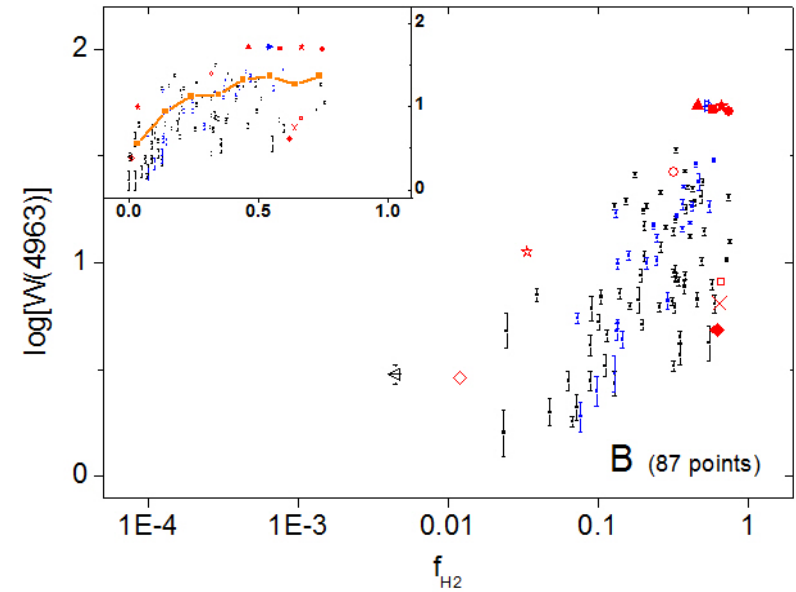
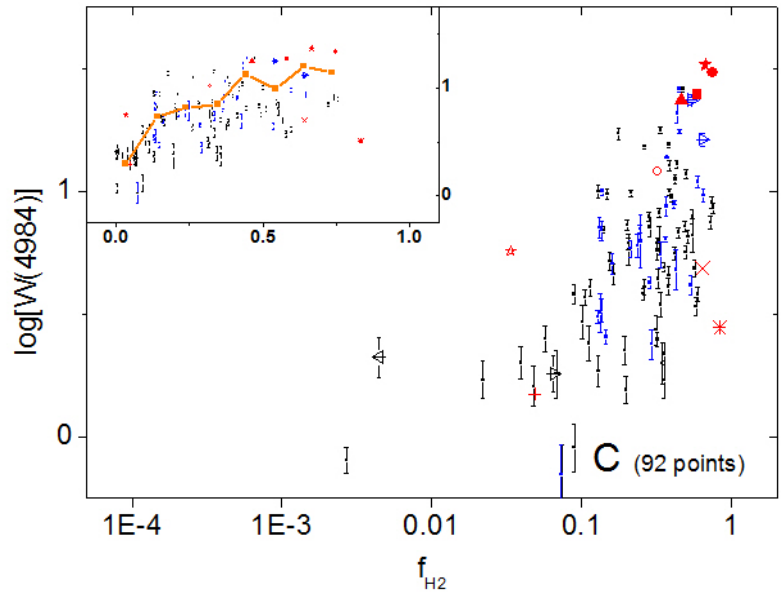
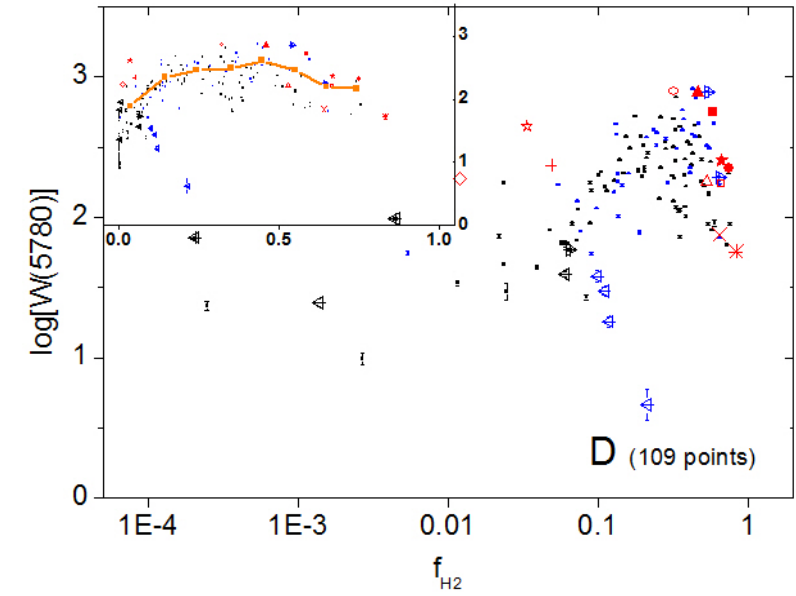

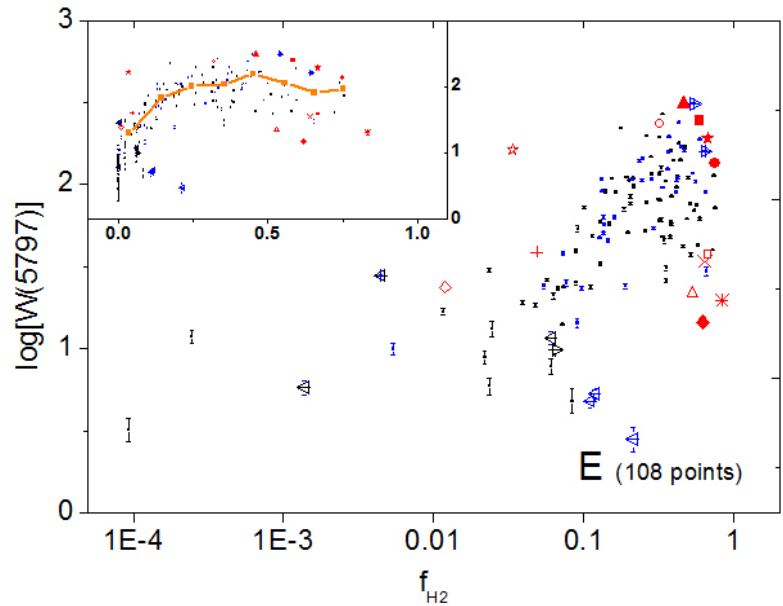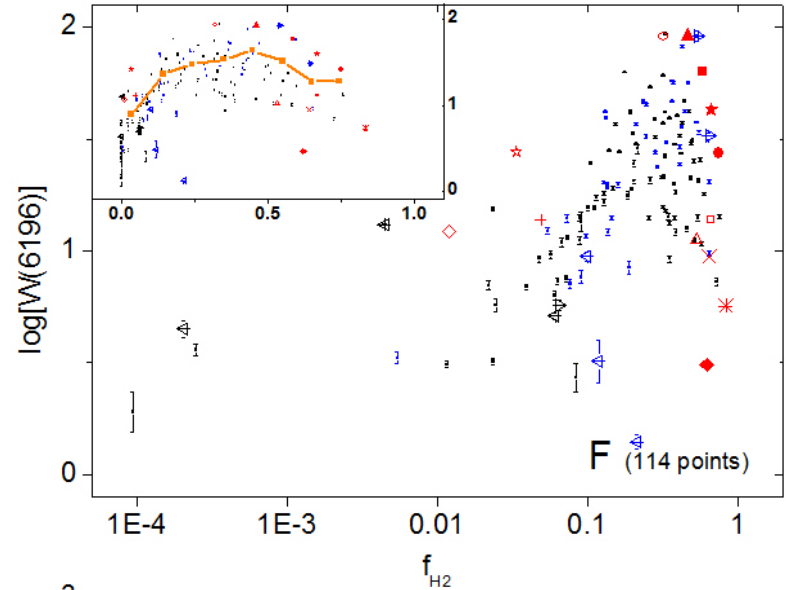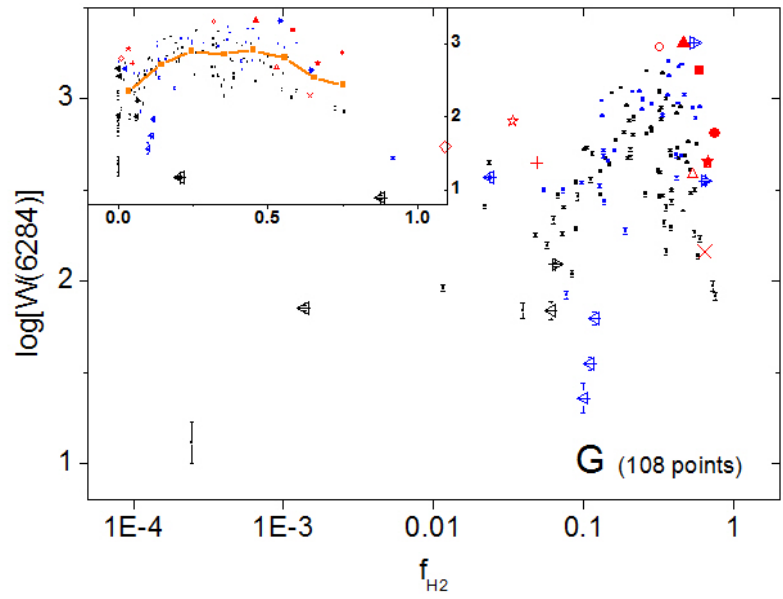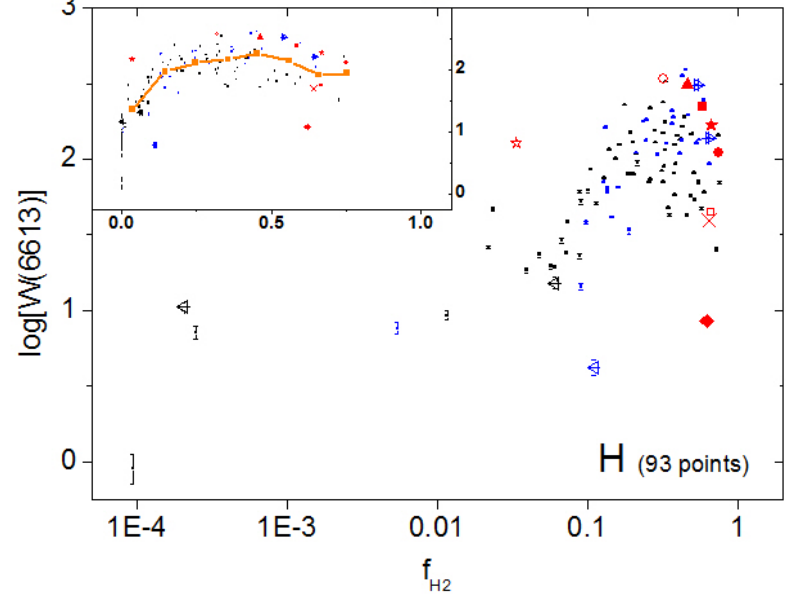

Figure B1. Behavior of *unnormalized* EWs of the 8 DIBs studied here as a function of $f_{H2}$ in log-log units, for direct comparison with Figure 4, which gives the EWs *normalized* to $E_{B-V}$. The panels are in the same order as in Figure 4. The pattern of Lambda-shaped behavior can still be seen even without the normalization to $E_{B-V}$, but the scatter is much larger compared to that in Figure 4 for $f_{H2} > 0.2$. Special stars are noted as follows: HD 147165 (Sig-Sco, ＋), HD 149757 (Zet-Oph, X), NGC2024-1 (●), HD 204827 (★), BD -14 5037 (■), Cyg OB2 #5 (▲), HD 37903 (△), HD 73882 (□), HD 37061 (◇), HD 62542 (◆), Herschel 36 (☆), HD 183143 (○), and HD 200775 (✳). The inset of each plot represents the same data with the abscissa plotted on a linear scale.

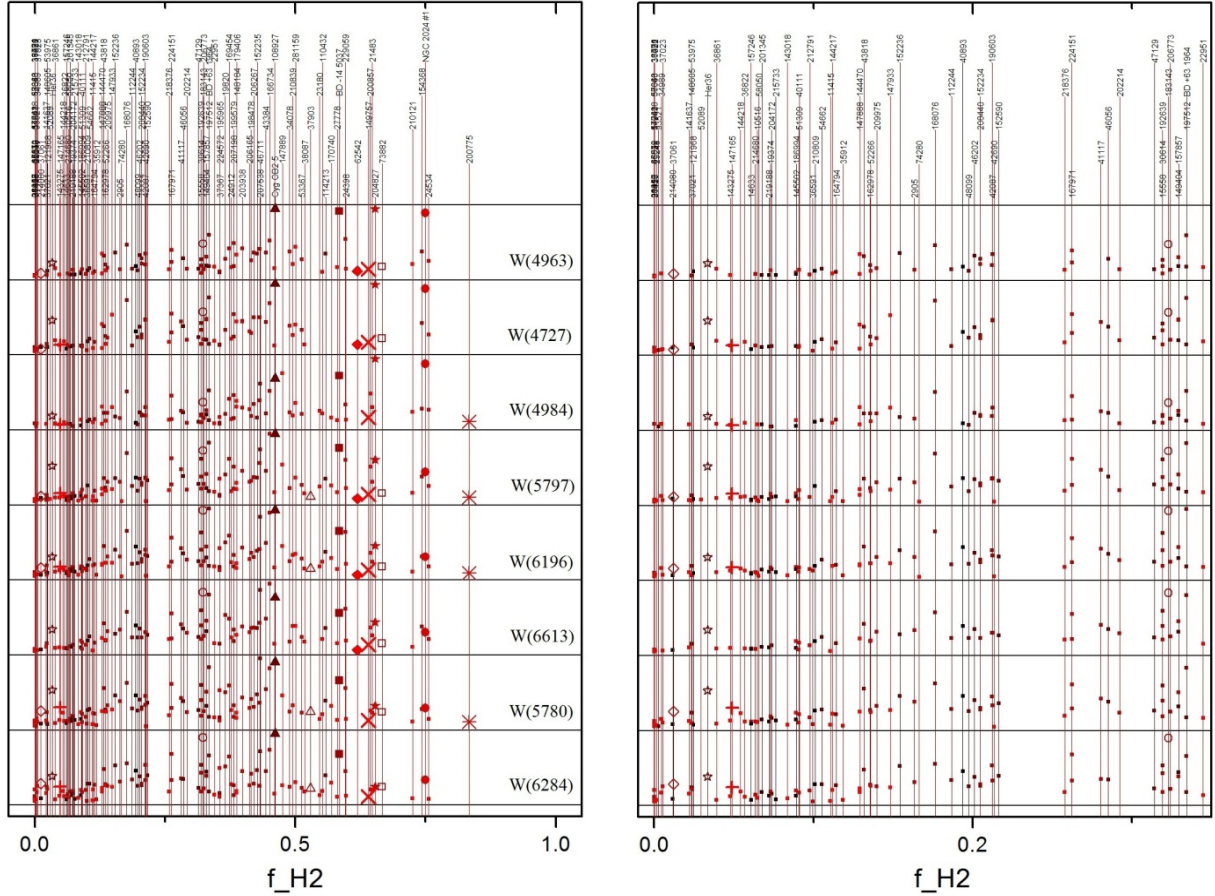

Figure B2. The equivalent widths of the eight DIBs studied here in the order of the sequence we proposed in Section 4.4.   The normalization to $E_{B-V}$ is not used.   The ordinate is 0 to 1 for each panel, representing the scaling of the DIBs to the highest values in each case, purely for presentation purposes.   The full scale of $f_{H2}$ from 0 to 1 is shown in the left half of the page, and the most crowded region, i.e., $0.0 < f_{H2} < 0.35$, is shown to the right.   The measurement corresponding to each star can be located by tracing the vertical line up to the star names shown on the top.   Figure B1 shows the global Lambda-shaped behavior.   Figure B2 can be used to examine the cosmic scatter between the stars of similar $f_{H2}$ values.